\documentstyle[12pt]{article}
\makeatletter
\def\eqnarray{%
\stepcounter{equation}%
\let\@currentlabel=\theequation
\global\@eqnswtrue
\global\@eqcnt\z@
\tabskip\@centering
\let\\=\@eqncr
$$\halign to \displaywidth\bgroup\@eqnsel\hskip\@centering
$\displaystyle\tabskip\z@{##}$&\global\@eqcnt\@ne
\hfil$\displaystyle{{}##{}}$\hfil
&\global\@eqcnt\tw@$\displaystyle\tabskip\z@{##}$\hfil
\tabskip\@centering&\llap{##}\tabskip\z@\cr }
\@addtoreset{equation}{section}
\makeatother
\def\theequation{\arabic{section}.\arabic{equation}}

\begin{document}
\vspace{0.3cm}
\boldmath
\begin{center}
{\Large{On Nature of Right-Handed Singlet in 331 Model and Anomalies before and after SSB}}
\end{center}
\unboldmath
\vspace{0.5cm}
\begin{center}
T. Kiyan, T. Maekawa, M. Masuda  and H. Taira\\ 
Department of Physics, Kumamoto University, Kumamoto 860-8555 Japan
\end{center}
\begin{abstract}
The gauge invariant and anomaly-free 331 model is studied with $R_\xi$ gauges in order to 
see which of the right-handed singlet between the case (i), assigning the bosonic singlet 
to the counterpart of $3$ and $\bar 3$ representations, and the case (ii), 
assigning fermionic siglet to the counterpart of the left-handed $3$ representation and 
assigning the antifermionic singlet to that of the left-handed $\bar 3$ representaion, 
should be assigned in physics but not mathematics to the counterparts of 
the left-handed fundamental $3$ representation and the left-handed complex conjugate 
$\bar 3$ representation. It is shown that through an effect on the Yukawa interaction of 
the transformation property of the singlet it depends on the choices of the case (i) or the case (ii) 
whether the anomaly coefficients even after the spontaneous symmetry breaking (SSB) vanish or not, 
and the coefficients vanish only in the case (ii) but do not in the case (i). Furthermore, 
it is pointed out that the BRS invariance after the SSB holds for the case (ii) and (i-b) of 
two possibilities in the choice for the mass eigenstates in the case (i) and thus the BRS invariance 
does not necessarily ensure renormalizability of the theory though the inverse of it holds.
\end{abstract}
\section{Introduction}\label{int}    @
Since the success of the standard model (SM) in physics below a few hundreds GEV, 
various extensions of the model based on the local gauge group are studied by many authors
\cite{1}--\cite{4} in order to understand some unanswered questions in the SM. 
Each of them is constructed covariantly on a local gauge group and then to be anomaly-free 
as well as renormalizability in the stage of the starting Lagrangian before the SSB. 
In particular, the 331 model is constructed on the gauge group 
${\rm SU(3)_C\otimes SU(3)_L\otimes U(1)_N}$ \cite{1} and has a property that the number 
of families is required to be a multiple of three to cancel anomalies\cite{5}. 
Anomalies do not cancel out within each family in contrast with the case of the SM 
but rather cancel out when contributions from all three families are taken into account for 
the quark sector. 

In the 331 model as well as in similar models, the representation $(1, 3, 0)$ is assigned to 
the left-handed leptons and $(1, {\it singlet}, y_N)$ to the right-handed ones in three families. 
On the other hand, the representation $(3,\bar 3, -1/3)$ is asigned to the left-handed quarks 
in the first and second families and the representation $(3, 3, 2/3)$ is assigned to 
the left-handed quarks in  the third family, while the representation $(3, {\it singlet}, y_N)$ 
is assigned to the right-handed counterparts. 
There is no problem on an interpretation of the ${\rm SU(3)_L}$ singlet in the lepton sector 
because only the fundamental representation is used for assignment and thus 
the right-handed singlet of the counterpart of the left-handed $3$ may be considered to be 
determined uniquely as the scalar under the fundamental representation. 
Also no problem occurs in the quark sector if only the fundamental 
or complex conjugate representation is used for assignment to the left-handed quarks 
because then the ``singlet" may be interpreted as the scalar under the assigned one of 
the representation. However, in the quark sectors in the model the representations $``3"$ and 
$``\bar 3"$ are assigned to the left-handed quarks. Then, there are several possibilities on 
an interpretation of the singlet. One of them is usually adopted as a mathematical scalar 
which means the scalar under both of the $3$ and $\bar 3$ representations or independent of 
the $3$ and $\bar 3$ representations\cite{1}. In the representation theory of ${\rm SU(3)}$, 
the scalar can be expressed in several configurations and one configuration can be rewritten 
into another one mathematically but each configuration should be considered to exhibit 
a physical object such as the color singlet in the quark model, i.e., bosonic singlet 
consisting of color and anticolor or fermionic singlet consisting of three colors 
(antifermionic singlet of three anticolors). The color singlet boson is different from 
the color singlet fermion physically but these configurations can be transformed into each other 
mathematically. The``singlet" in the above assignment may be apriori any one among 
a bosonic singlet, fermionic or antifermionic singlet because the singlet is not a composite 
and may be a ``scalar" independent of the content. It is, however, important for us to determine 
the kind of the singlet because it has an effect on the form of the Yukawa interaction\cite{6} 
whether the bosonic or fermionic (antifermionic) singlet is adopted and further it depends on 
the nature of the singlet whether the triangle anomaly coefficients\cite{5} disappear or not 
after the SSB\cite{7}.

In a previous article\cite{6}, it is pointed out in the case of the $331$ model without 
gauge fixing that even though the starting Lagrangian has gauge invariant and thus renormalizable 
form the renormalizability after the SSB is not necessarily guaranteed if the content of 
the singlet assigned to the right-handed quarks is not taken into account. This is because 
in the $331$ model the representations $(3, 3, y_N)$ and $(3, {\bar 3},y^\prime_N)$ of  
${\rm SU(3)_C \otimes SU(3)_L \otimes U(1)_N}$ are assigned to the left-handed quarks 
in the families, while we have two possibilities for the assignment of the representations to 
the right-handed quarks, i.e., (i) $(3, 1, y_N)$ for all right-handed quarks, and 
(ii) $(3, 1_3,y_N)$ and $(3, 1_{\bar 3}, y_N)$ for the fermionic and antifermionic singlets. 
The case (i) means that the $``1"$ of ${\rm SU(3)_L}$ is independent of whether this state is 
accompanied with $3$ or $\bar 3$ of ${\rm SU(3)_L}$ and many authors\cite{1} except for us adopt 
this possibility as far as we know. The case (ii) means that there exist the fermionic 
and antifermionic singlets, where $``1_3"$ is only the scalar with respect to 
the transformation of the fundamental triplet and $``1_{\bar 3}"$ is only a scalar with 
respect to the transformation of the complex conjugate representation corresponding to 
the fermion and antifermion with respect to the color quantum number in the quark model, respectively. 
These two cases bring about no different results except for the Yukawa interaction of 
the quarks with the scalar fields but it is definitely important for us to distinguish 
these possibilities in order to construct the Yukawa interactions in the starting stage of 
the weak interaction bases. Through the Yukawa interaction, the quark fields get their masses after 
the SSB\cite{7} and the mass eigenstates (physical fields) are given in terms of 
the linear combination of the weak interaction bases (fields with superscript $0$). If the case (i) 
is adopted, the mass eigenstates through the interactions need the mixing of 
the left-handed quarks belonging to the different representations in order to avoid 
a flavor change (the case (i-a)) or no-mixing to forgive a flavor change through 
an interaction with vacuum (the case (i-b)). These mixing or no-mixing brings about nonvanishing 
effect of the triangle anomalies through the interactions of the quark current with 
the gauge bosons and in the former case the flavor changing neutral current (FCNC)\cite{8} appears. 
But if the case (ii) is adopted, the above mixing is not needed and thus the triangle anomalies 
as well as the FCNC do not appear\cite{6}. 

In this article, the $331$ model is discussed in the cases (i) and (ii) with the $R_\xi$ 
gauges\cite{9} as well as the ghost terms\cite{10}. In addition, the BRS transformation\cite{11} 
after the SSB is examined. In \S\ref{for}, the notations and Lagrangian are introduced. 
In \S\ref{ssb}, the SSB is considered and the Lagrangians are expressed in terms of 
the mass eigenstates. In \S\ref{gau}, the gauge fixing to $R_\xi$ and ghost Lagrangians are given 
in some detail because almost authors treat with the model by using the Higgs mechanism. 
In \S\ref{pro}, the propagators for the gauge particles, scalars, fermions and ghosts 
are discussed. In \S\ref{fer}, the fermion currents are expressed in terms of the mass eigenstates 
and in \S\ref{tri}, the fermion triangles anomalics are discussed and it is pointed that 
the anomaly coefficients disappear only in the case (ii) for the singlet. 
\S\ref{brs} is devoted to a brief discussion of the BRS trnsformation and it is seen that 
there exists a case where the BRS invariance holds even in the case of non-zero anomalies.
\section{Formalism}\label{for}
In this section, notations and a brief explanation will be given for the $331$ model\cite{1}. 
An assignment of the representations to the basic quarks (fermions) is adopted as follows:
\begin{eqnarray}
&& 
l^0_{aL}=\left(
\begin{array}{c}
\nu^0_a \\
e^0_a \\
E^0_a
\end{array}
\right)_L\sim(1,3,0),\ \ \nonumber\\
&&
\nu^0_{aR}\sim (1,1,0),\ \ \ 
e^0_{aR}\sim (1,1,-1),\ \ \ 
E^0_{aR}\sim (1,1,1),\nonumber\\
&& 
Q^{0}_{iL}=\left( 
\begin{array}{c}
d^0_i \\
u^0_i \\
J^0_i
\end{array}
\right)_L\sim(3,{\bar 3},-1/3),\ \ \label{21}\\\
&& 
u^0_{iR}\sim(3,1(1_{\bar 3}),2/3),\ \ \ 
d^0_{iR}\sim(3,1(1_{\bar 3}),-1/3),\ \ \ 
J^0_{iR}\sim(3,1(1_{\bar 3}),-4/3),\nonumber\\
&& 
Q^{0}_{3L}=\left( 
\begin{array}{c}
u^0_3 \\
d^0_3 \\
J^0_3
\end{array}
\right)_L\sim(3,3,2/3),\ \ \nonumber\\
&& 
u^0_{3R}\sim(3,1(1_3),2/3),\ \ \ 
d^0_{3R}\sim(3,1(1_3),-1/3),\ \ \ 
J^0_{3R}\sim(3,1(1_3),5/3),\nonumber
\end{eqnarray}
where $a$ takes the values $1,2,3$ and $i$ takes $1,2$. 
The singlet term for the lepton should be written as $1_3$ in our convention but it is not necessary 
to distinguish the notations for the leptons because only fundamental triplet representation 
is used for the left-handed leptons and in these cases the singlet terms are understood naturally 
as $1_3$. However, for the quarks, both fundamental and complex conjugate triplet representations 
for the left-handed quarks are used and we should distinguish a transformation property of 
the singlet whether the singlet is accompanied with the fundamental or complex conjugate triplet 
representation. As pointed out in a previous article\cite{6} the singlet of ${\rm SU(3)}$ 
appears through three ways. One of them consists of the basic quark and antiquark, and 
the other consists of the three quarks (or antiquarks). Of course, their configurations are 
mathematically (or group theoretically) unique in the sense that one configuration of them can be 
transformed into the other. However, they can be distinguished physically as the bosonic, 
fermionic and antifermionic fields though the singlet in the assignment is not composite. 

In order to give the quarks mass through the SSB, three scalar fields are introduced
\begin{eqnarray}
&&
\chi=
\left( 
\begin{array}{c}
\chi^- \\
\chi^{--} \\
\chi^0
\end{array}
\right)_L\sim(1,3,-1),\\
&& 
\rho=
\left( 
\begin{array}{c}
\rho^+ \\
\rho^0 \\
\rho^{++}
\end{array}
\right)_L\sim(1,3,1),\nonumber\\ 
&&
\eta=
\left( 
\begin{array}{c}
\eta^0 \\
\eta^- \\
\eta^+
\end{array}
\right)_L\sim(1,3,0).\nonumber
\end{eqnarray}
It is noted that the representations $\chi$, $\rho$ and $\eta$ may be considered as 
the irreducible components of a reducible field $\Phi$ of 
${\rm SU(3)_C\otimes SU(3)_L\otimes U(1)_N}$. Then, the notation $\Phi=(\chi, \rho, \eta )^T$ 
will be used.

The charge operator $Q$ is given by $Q=T^3-\sqrt 3 T^8 +N$, where $T^3$ and $T^8$ together with 
$T^i(i=1\sim 8)$ are the generators of ${\rm SU(3)_{L}}$ and have the representation matrix 
$\lambda^i/2$ for the $3$ representation and $-\lambda^T/2$ for the $\bar 3$ representation of 
${\rm SU(3)_L}$. The value $y_n$ of the generator $N$ of ${\rm U(1)}$ is chosen to give 
the charge on the particle. 

It follows that the quantities consisting of a linear combination of 
the corresponding leptons in the three families are subject to the same transformation as 
that of the original ones under ${\rm SU(3)_C\otimes SU(3)_L\otimes U(1)_N}$, e.g.,
\begin{eqnarray}
\left( 
\begin{array}{c} 
(U^\nu \nu^0)_a \\ (U^e e^0)_a \\ (U^J E^0)_a 
\end{array} 
\right)_L \sim (1,3, 0),
\ \ \ \ \ \ \ \ \ \ \ \ 
(U^\nu \nu^0)_{aR}\sim (1,1,0),\ a=1,2,3,\label{23}
\end{eqnarray}
where the $U^\nu$ etc. denote some $3\times 3$ unitary matrices. For the quarks, 
on the other hand, the situation is somewhat different from that for leptons. The state of 
a linear combination of the right-handed quarks in the three families has the same transformation 
under the gauge group as the original ones such as $(Uu^0)_{aR}\sim (3, 1, 2/3)$ in the case (i). 
In the case (ii),e.g., $(Uu^0)_{aR}\not\sim(3, 1_3 $\rm or$ 1_{\bar 3}, 2/3)$ 
because the singlet $``1_3"$ is a scalar under the $3$ representation of $SU(3)_L$ but not under 
the $\bar 3$ representation and the singlet $``1_{\bar 3}"$ is also, while the quantity 
$(Uu^0)_{aR}$ means the linear combination of the quarks in the three families. 
For the left-handed quarks,e.g.,
\begin{eqnarray}
\left( 
\begin{array}{c} (U_{1} u^0)_3 \\ (U_{2} d^0)_3 \\ J^0_3 
\end{array} 
\right)_L \not \sim (3,3, 2/3), 
\ \ \ \ \ \ 
\left( 
\begin{array}{c} (U_{3} u^0)_i \\ (U_{4} d^0)_i \\ (U^JJ^0)_i 
\end{array} 
\right)_L \not \sim (3,{\bar 3}, -1/3), \ (i=1,2,) \label{24}
\end{eqnarray}
where $U_{1}$ etc. denote some $3\times 3$ unitary matrices except for unit matrix, 
and $U^J$ $2\times 2$ matrix depending on the first and second families and it is noted 
that $J^0_3$ does not mix with $J^0_1$ and $J^0_2$ because of the different charges. 
On the other hand, the followings hold in the case (ii)
\begin{eqnarray}
(U^uu^0)_{iR}\sim(3, 1_{\bar 3}, 2/3),
\ \ \ \ \ \ \ \ \ \ 
\left( 
\begin{array}{c} (U^u u^0)_i \\ (U^d d^0)_i \\ (U^JJ^0)_i 
\end{array} 
\right)_L\sim(3,{\bar 3}, -1/3), \ (i=1,2)\label{25}
\end{eqnarray}
where $U^{u}$ etc. denote some $2\times 2$ matrix on the quarks in the first and 
the second families. It is noted that the relations $(\ref{24})$ mean an impossibility of 
introducing the BRS transformation\cite{11} in contrast to the second of (\ref{25}).

In the above assignment of the fermions, the leptons in the three families can be mixed with 
the corresponding leptons because common representation $3$ or $1\ (\equiv 1_3)$ is adopted, 
while the quarks in the first and second families should not be mixed with the corresponding ones 
in the third family because they are subject to the different representations and have 
the different $y_n$ charges. 

In what follows, the color symmetry is not treated 
and thus is omitted unless stated otherwise. The gauge covariant kinetic energy Lagrangians 
for the fermions, scalars and gauge fields are given as follows:
\begin{eqnarray}
&& {\cal L}_f= \sum_{a=1}^3
\left\{
{\bar{l}}^0_{aL}i{\cal D}_\mu\gamma^\mu{l}^0_{aL}
+{\bar \nu}^0_{aR}i{\cal D}_\mu\gamma^\mu{\nu}^0_{aR}
+{\bar e}^0_{aL}i{\cal D}_\mu \gamma^\mu {e}^0_{aL}
+{\bar E}^0_{aR}i{\cal D}_\mu \gamma^\mu {E}^0_{aR}
\right.\nonumber\\
&& \hspace{1cm} 
+\left.
{\bar Q}^0_{aL}i{\cal D}_\mu\gamma^\mu{Q}^0_{aL}
+{\bar u}^0_{aR}i{\cal D}_\mu \gamma^\mu {u}^0_{aR}
+{\bar d}^0_{aR}i{\cal D}_\mu \gamma^\mu {d}^0_{aR}
+{\bar J}^0_{aR}i{\cal D}_\mu \gamma^\mu {J}^0_{aR}
\right\},\label{26a}\\
&& 
{\cal L}_{sc}=({\cal D}^\mu\chi)^\dag({\cal D}_\mu\chi)
+({\cal D}^\mu \rho)^\dag ({\cal D}_\mu \rho)
+({\cal D}^\mu \eta)^\dag ({\cal D}_\mu \eta)
-V(\chi, \ \rho,\ \eta) ,\label{26b}\\
&& {\cal L}_V= -\frac {1}{4}F^j_{\mu \nu}{F^j}^{\mu\nu}
-\frac{1}{4}B^{\mu\nu}B_{\mu\nu},\label{26c} 
\end{eqnarray}
where the covariant derivatives for each field are given 
without using the different notations as follows:
\begin{eqnarray} 
&&
{\cal D}_\mu Q^0_{iL}=
(\partial_\mu+i\frac{g}{2}\lambda^T\cdot A_\mu+ig_N\frac{1}{3}B_\mu)
Q^0_{iL},\nonumber\\
&&
{\cal D}_\mu Q^0_{3L}=
(\partial_\mu-i\frac{g}{2}\lambda\cdot A_\mu-ig_N\frac{2}{3}B_\mu)
Q^0_{3L},\nonumber\\
&&
{\cal D}_\mu u^0_{aR}=
(\partial_\mu-ig_N\frac{2}{3}B_\mu)
u^0_{aR},\nonumber\\
&&
{\cal D}_\mu d^0_{aR}=
(\partial_\mu+ig_N\frac{1}{3}B_\mu)
d^0_{aR}, \label{27}\\ 
&&
{\cal D}_\mu J^0_{iR}=
(\partial_\mu+ig_N\frac{4}{3}B_\mu)
J^0_{iR},\nonumber\\
&&
{\cal D}_\mu J^0_{3R}=
(\partial_\mu-ig_N\frac{5}{3}B_\mu)
J^0_{3R},\nonumber\\
&&{\cal D}_\mu \varphi=
(\partial_\mu-i\frac{g}{2}\lambda\cdot A_\mu-ig_Ny_\varphi B_\mu)
\varphi,\nonumber
\end{eqnarray}
with $y_\varphi=-1$ for $\varphi=\chi$, 
$y_\varphi=1$ for $\varphi=\rho$ and 
$y_\varphi=0$ for $\varphi=\eta$.

The field strengths $F^j_{\mu\nu}$ and $B_{\mu\nu}$ are
\begin{eqnarray}
&& F^j_{\mu \nu }=\partial_\mu A^j_\nu-\partial_\nu A^j_\mu 
+gc_{jkl}A^k_\mu A^l_{\nu}, \label{28}\\
&& B_{\mu \nu}=\partial_\mu B_\nu - \partial_\nu B_\mu,\nonumber
\end{eqnarray}
where $c_{ijk}\ (i,j,k=1,2,\cdots,8)$ are the structure constants of ${\rm SU(3)}$.

The interaction potential of the scalar fields and 
the Yukawa interactions of the fermions with the scalar fields 
are given as follows:  
\begin{eqnarray}
V(\chi,\ \rho,\ \eta )
&&
=c_1{\left(\chi^\dag\chi-\frac{1}{2}\chi_v^2\right)}^2
+c_2{\left(\rho^\dag\rho-\frac{1}{2}\rho_v^2\right)}^2
+c_3{\left(\eta^\dag\eta-\frac{1}{2}\eta_v^2\right)}^2\nonumber\\
&&
+c_4(\chi^\dag\chi-\chi_v^2)(\rho^\dag\rho-\rho_v^2 )
+c_5(\chi^\dag\chi-\chi_v^2)(\eta^\dag\eta-\eta_v^2 )
+c_6(\rho^\dag\rho-\rho_v^2)(\eta^\dag\eta-\eta_v^2 )\nonumber\\
&&
+c_7(\chi^\dag\rho)(\chi^\dag\rho)^\dag
+c_8(\chi^\dag\eta)(\chi^\dag\eta)^\dag
+c_9(\rho^\dag\eta)(\rho^\dag\eta)^\dag,\label{29}
\end{eqnarray}
\begin{eqnarray}
{\cal L}_Y 
&&
={(\overline{ l^0_1},\ \overline{l^0_2},\ \overline{ l^0_3})}_L\mu^\chi 
\left( 
\begin{array}{c} 
E^0_1 \\ 
E^0_2 \\ 
E^0_3 
\end{array}
\right)_R\chi
%
+{(\overline{l^0_1},\ \overline{ l^0_2},\ \overline{l^0_3})}_L\mu^\rho 
\left( 
\begin{array}{c} 
e^0_1 \\ 
e^0_2 \\ 
e^0_3 
\end{array}
\right)_R\rho
%
+{(\overline{l^0_1},\ \overline{ l^0_2},\ \overline{l^0_3})}_L\mu^\eta 
\left( 
\begin{array}{c} 
\nu^0_1 \\ 
\nu^0_2 \\ 
\nu^0_3 
\end{array}
\right)_R\eta\nonumber\\
&&
+\Gamma^\chi_3\overline{Q^0_{3L}}J^0_{3R}\chi
+{(\overline{Q^0_1},\overline{Q^0_2})}_L
\left( 
\begin{array}{cc}
\Gamma^\chi_{11} & \Gamma^\chi_{12}\\ 
\Gamma^\chi_{21} & \Gamma^\chi_{22}
\end{array}
\right)
\left( 
\begin{array}{c} 
J^0_1 \\ 
J^0_2 
\end{array}
\right)_R \chi^*\nonumber\\
&&
+\overline{Q^0_{3L}}(\Gamma^\rho_{31},\Gamma^\rho_{32},\Gamma^\rho_{33})
\left( 
\begin{array}{c} 
d^0_1\\ 
d^0_2\\ 
d^0_3
\end{array}
\right)_R\rho
+{(\overline{Q^0_1},\overline{ Q^0_2})}_L
\left( 
\begin{array}{ccc}
\Gamma^\rho_{11} & \Gamma^\rho_{12}& \Gamma^\rho_{13} \\ 
\Gamma^\rho_{21} & \Gamma^\rho_{22}& \Gamma^\rho_{23}
\end{array}
\right)
\left( 
\begin{array}{c} 
u^0_1 \\ 
u^0_2 \\ 
u^0_3 
\end{array}
\right)_R \rho^*\nonumber\\
&&
+\overline{Q^0_{3L}}(\Gamma^\eta_{31},\Gamma^\eta_{32},\Gamma^\eta_{33})
\left( 
\begin{array}{c} 
u^0_1 \\ 
u^0_2 \\ 
u^0_3 
\end{array}
\right)_R\eta
+{(\overline{Q^0_1},\overline{Q^0_2})}_L
\left( 
\begin{array}{ccc}
\Gamma^\eta_{11} & \Gamma^\eta_{12}& \Gamma^\eta_{13}\\ 
\Gamma^\eta_{21} & \Gamma^\eta_{22} & \Gamma^\eta_{23}
\end{array}
\right)
\left( 
\begin{array}{c} 
d^0_1 \\ 
d^0_2 \\ 
d^0_3 
\end{array}
\right)_R \eta^*\nonumber\\
&&
+h.c.,\label{210}
\end{eqnarray}
where $\mu^\varphi(\varphi=\chi,\ \rho,\ \eta)$ is $3\times 3$ constant matrix. 
${\cal L}_Y$ in $(\ref{210})$ is written in the case of the assignment of the bosonic singlet 
$(``1")$ of ${\rm SU(3)_L}$ to the all right-handed quarks. Of course, the expressions for 
the interaction between the left-handed and the exotic right-handed quarks with the scalar $\chi$ 
are the same independent of the nature of the singlets and the term involving the quarks 
in the first and second families is not mixed with the corresponding one in the third family 
because the charge on the exotic quarks in the first and second families differs from that on 
the exotic quark in the third family. However, as is seen from $(\ref{210})$ the interactions of 
the left-handed and the right-handed quarks with $\rho$ or $\eta$ in the three families will 
give rise to a mixture of the quarks in the three families to give masses to the quarks. 
In the case (ii) of taking into account the nature of the singlet, i.e., fermionic and 
antifermionic singlets, the interactions of the quarks in the first and second families through 
$\rho$ or $\eta$ are not mixed with those in the third family as in the case for 
the exotic right-handed quarks through $\chi$ because in this case the interactions of the $u$ 
and $d$ quarks with the scalar bosons are given by putting 
$\Gamma^{\rho,\eta}_{3i}=\Gamma^{\rho,\eta}_{i3}=0$ for $i=1,2$ in $(\ref{210})$ and thus 
the left-handed quarks in the first and second families interact with the right-handed quarks 
through $\rho^*$ or $\eta^*$, while those in the third family interact through $\rho$ or $\eta$ 
such as those in the case of $\chi$. These results affect the physical fields (mass eigenstates) 
of the quarks after the SSB and then the interactions between the quark current and 
the gauge bosons which play an important role for the discussion of the triangle anomalies\cite{5}.   

The Lagrangian ${\cal L}_f$ in $(\ref{26a})$ can be rewritten in terms of 
the fermion currents\cite{12} as follows
\begin{eqnarray}
{\cal L}_f
&&
=\overline{l^0_{aL}}i\gamma^{\mu}\partial_{\mu}{l^0_{aL}}
+\overline{\nu^0_{aR}}i\gamma^{\mu}\partial_{\mu}{\nu^0_{aR}}
+\overline {e^0_{aR}}i\gamma^{\mu}\partial_{\mu}{e^0_{aR}}
+\overline {E^0_{aR}}i\gamma^{\mu}\partial_{\mu}{E^0_{aR}}\nonumber\\
&&
+\overline{Q^0_{aL}}i\gamma^{\mu}\partial_{\mu}{Q}^0_{aL}
+\overline{u^0_{aR}}i\gamma^{\mu}\partial_{\mu}{u}^0_{aR}
+\overline{d^0_{aR}}i\gamma^{\mu}\partial_{\mu}{d}^0_{aR}
+\overline{J^0_{aR}}i\gamma^{\mu}\partial_{\mu}{J}^0_{aR}\nonumber\\
&&
+\frac{g}{2\sqrt 2}
\left[
J^\mu_WW^+_\mu+J^\mu_XX^+_\mu+J^\mu_YY^{++}_\mu+h.c.
\right]\nonumber\\
&&
+eJ^\mu_{em}A_\mu
+\frac{g}{2c_W}J^\mu_ZZ_\mu
+\frac{g}{2\sqrt{1-3t^2_W}}J^\mu_{Z^\prime}Z^\prime_{\mu},\label{211}
\end{eqnarray}
where 
\begin{eqnarray*}
&&
J^\mu_{em}
=\sum_{a}
{\overline{l^0}_{aL}}\gamma^\mu Ql^0_{aL}
-\sum_{a}
{\overline{e^0}_{aR}}\gamma^\mu e^0_{aR}
+\sum_{a}
{\overline{E^0}_{aR}}\gamma^\mu E^0_{aR}
+\sum_{a}
{\overline{Q^0}_{aL}}\gamma^\mu QQ^0_{aL}\\
&&\hspace{0.8cm}
+\frac{2}{3}\sum_{a}
{\overline{u^0}_{aR}}\gamma^\mu u^0_{aR}
-\frac{1}{3}\sum_{a}
{\overline{d^0}_{aR}}\gamma^\mu d^0_{aR}
-\frac{4}{3}\sum_{j}
{\overline{J^0}_{jR}}\gamma^\mu J^0_{jR}
+\frac{5}{3}{\overline{J^0}_{3R}}\gamma^\mu J^0_{3R},\\
&&
J^\mu_W
=\sum_{a}
{\overline{l^0}_{aL}}\gamma^\mu(\lambda_4-i\lambda_5)l^0_{aL}
+{\overline{Q^0}_{3L}}\gamma^\mu(\lambda_1+i\lambda_2)Q^0_{3L}
+\sum_{j}
{\overline{Q^0}_{jL}}\gamma^\mu(-\lambda^T_1-i\lambda^T_2)Q^0_{jL},\\
&&
J^\mu_X
=\sum_{a}
{\overline{l^0}_{aL}}\gamma^\mu(\lambda_4-i\lambda_5)l^0_{aL}
+{\overline{Q^0}_{3L}}\gamma^\mu(\lambda_4-i\lambda_5)Q^0_{3L}
+\sum_{j}
{\overline{Q^0}_{jL}}\gamma^\mu(-\lambda^T_4+i\lambda^T_5)Q^0_{jL},\\
&&
J^\mu_Y
=\sum_{a}
{\overline{l^0}_{aL}}\gamma^\mu(\lambda_6-i\lambda_7)l^0_{aL}
+{\overline{Q^0}_{3L}}\gamma^\mu(\lambda_6-i\lambda_7)Q^0_{3L}
+\sum_{j}
{\overline{Q^0}_{jL}}\gamma^\mu(-\lambda^T_6+i\lambda^T_7)Q^0_{jL},\\
&&
J^\mu_Z
=\sum_{a}
{\overline{l^0}_{aL}}\gamma^\mu(\lambda_3)l^0_{aL}
+{\overline{Q^0}_{3L}}\gamma^\mu\lambda_3Q^0_{3L}
+\sum_{j}
{\overline{Q^0}_{jL}}\gamma^\mu(-\lambda^T_3)Q^0_{jL}-2s^2_WJ^\mu_{em},\\
&&
J^\mu_{Z^\prime}
=\sum_{a}
{\overline{l^0}_{aL}}\gamma^\mu(\lambda_8-\sqrt 3t^2_W\lambda_3)l^0_{aL}
+{\overline{Q^0}_{3L}}\gamma^\mu(\lambda_8-\sqrt 3t^2_W \lambda_3)Q^0_{3L}\\
&&\hspace{0.8cm}
+\sum_{j}
{\overline{Q^0}_{jL}}\gamma^\mu(-\lambda^T_8+\sqrt 3t^2_W\lambda^T_3)Q^0_{jL}
+2\sqrt{3}t^2_WJ^\mu_{em},
\end{eqnarray*}
\begin{eqnarray*}
&&
\sqrt{2}W^\pm_\mu =A^1_\mu \mp iA^2_\mu ,\ \ 
\sqrt{2}X^\pm_\mu =A^4_\mu \pm iA^5_\mu,\ \ 
\sqrt{2}Y^{\pm\pm}_\mu =A^6_\mu \pm iA^7_\mu,\\
&&
A_\mu=s_WA^3_\mu
+c_W(-\sqrt 3t_W A^8_\mu
+\sqrt{1-3t^2_W}B_\mu),\\
&&
Z_\mu=c_W A^3_\mu-s_W(-\sqrt{3}t_W A^8_\mu+\sqrt{1-3t^2_W}B_\mu),\\
&&
Z^\prime_\mu=\sqrt{1-3t^2_W}A^8_\mu+\sqrt{3}t_WB_\mu,\\
&&
t_W\equiv\tan\theta_W=\frac{g_N}{\sqrt{g^2+3g^2_N}},
\ \ \ \ 
c_W=\cos\theta_W,
\ \ \ \ 
e=\frac{gg_Y}{g^2+g_Y^2}.
\end{eqnarray*}
The summation convention will be used here and hereafter over $a=1,2,3$ and $j=1,2$. 
It is noted that though the following is obvious from the assignment of the representations to 
the basic quarks, the above currents for the quark part have the structure that the quarks 
in the first and second families are not mixed with those in the third family. It is, thus, 
expected that the expressions for the currents after the SSB should have the form similar to 
those for the above weak interaction bases in order to make the theory renormalizable, 
otherwise the vanishing conditions (trace condition in terms of the representation matrices) for 
the triangle anomalies for various processes\cite{5} will not be expressed in terms of 
the representation matrices and thus the vanishing condition for all possible processes will not be 
satisfied in general. It is shown that this result depends on whether the nature of the singlet 
is taken into account or not\cite{6}.
\section{SSB and particle masses}\label{ssb}

By introducing the $VEV (\chi_v)$ for $\chi$ and writing,
\begin{eqnarray}
\chi\equiv\frac{1}{\sqrt{2}}\chi_v
+\frac{1}{\sqrt{2}}\chi_h
=\frac{1}{\sqrt{2}}
\left( 
\begin{array}{c} 
0 \\ 
0 \\ 
\chi_v 
\end{array} 
\right)
+\frac{1}{\sqrt{2}}
\left( 
\begin{array}{c} 
\chi^-_h \\ 
\chi^{--}_h 
\\ \chi^0_h  
\end{array} 
\right),\label{31}
\end{eqnarray}
the symmetry ${\rm SU(3)_C\otimes SU(3)_L\otimes U(1)_N}$ is broken down to 
${\rm SU(3)_C\otimes SU(2)_L\otimes U(1)_Y}$, with the hypercharge operator $Y=N -{\sqrt 3}T_8$. 
It is noted that although the group is broken down to ${\rm SU(3)_C\otimes SU(2)_L\otimes U(1)_Y}$ 
and the 2 and ${\bar 2}$ representations of ${\rm SU(2)}$ are equivalent, the contents of 
the quarks restricted to the ${\rm SU(3)_C\otimes SU(2)_L\otimes U(1)_Y}$ in the families are 
different from those in the standard model (SM). For example, the transformation of the doublets 
in the case of 
the SM is as follows
\begin{eqnarray}
\left( 
\begin{array}{c} 
u_a \\ 
d_a 
\end{array} 
\right)_L \sim (3, 2,1/6),\ a=1,2,3,\label{32}
\end{eqnarray}
while the corresponding doublets except for the $J$ quarks are now
\begin{eqnarray}
&&
\left( 
\begin{array}{c} 
d_i \\ 
u_i 
\end{array} 
\right)_L \sim (3, {\bar 2},1/6),\ i=1,2,\hspace{1.5cm}
\left( 
\begin{array}{c} 
u_3 \\ 
d_3 
\end{array} 
\right)_L \sim (3, 2,1/6).\label{33} 
\end{eqnarray}
The first doulets with (${\bar 2}$) in $(\ref{33})$ may be rewritten in those with 
the $2$ or ${\bar 2}$ representation as follows
\begin{eqnarray}
&&
\left( 
\begin{array}{c} 
d^\dag _i \\ 
u^\dag_i 
\end{array} 
\right)_L \sim(\bar 3,  2,-1/6),\hspace{1.5cm}
\left( 
\begin{array}{c} 
u^\dag _i \\ 
-d^\dag_i 
\end{array} 
\right)_L \sim (\bar 3, \bar 2,-1/6),\ i=1,2.\label{34}
\end{eqnarray}
The $2$ and ${\bar 2}$ representations of ${\rm SU(2)}$ are equivalent as is well known but 
it should be taken care of that the doublets transforming according to them are related to 
each other in the complex conjugate as above and they are not the same. 

In a similar fashion, by the $VEV$ for $\rho$ and writing it
\begin{eqnarray}
\rho\equiv\frac{1}{\sqrt 2}\rho_v
+\frac{1}{\sqrt{2}}\rho_h
=\frac{1}{\sqrt{2}}
\left( 
\begin{array}{c} 
0 \\ 
\rho_v \\ 
0 
\end{array} 
\right)
+\frac{1}{\sqrt{2}}
\left( 
\begin{array}{c} 
\rho^+_h \\ 
\rho^0_h \\ 
\rho^{++}_h  
\end{array} 
\right),\label{35}
\end{eqnarray}
the symmetry ${\rm SU(2)_L\otimes U(1)_Y}$ is broken down to ${\rm U(1)_{em}}$. 
In order to give mass to all quarks, the $VEV$ for $\eta$ is needed 
and $\eta$ is written as follows:
\begin{eqnarray}
\eta\equiv\frac{1}{\sqrt{2}}\eta_v
+\frac{1}{\sqrt{2}}\eta_h
=\frac{1}{\sqrt{2}}
\left( 
\begin{array}{c} 
\eta_v \\ 
0 \\ 
0 
\end{array} 
\right)
+\frac{1}{\sqrt{2}}
\left( 
\begin{array}{c} 
\eta^0_h \\ 
\eta^-_h \\ 
\eta^+_h 
\end{array} 
\right).\label{36}
\end{eqnarray}
where here and hereafter the same notations for the component of the $VEV$ of the scalars and the $VEV$ 
itself are used for confusion will not occur. It is noted that the ${\rm U(1)_{em}}$ symmetry 
is not affected by introducing the $VEV$ for $\eta$.

The scalar potential $(\ref{29})$, from which the mass of the scalar particles is determined, is 
rewritten in terms of the fields with the subscript $h$ by the replacement of the following forms
\begin{eqnarray}
&&
\chi^\dag\chi
\rightarrow
\frac{1}{2}(\chi^\dag_h+\chi^\dag_v)(\chi_h+\chi_v),\nonumber\\
&&
\chi^\dag\rho
\rightarrow
\frac{1}{2}(\chi^\dag_h+\chi^\dag_v)(\rho_h+\rho_v).\label{37}
\end{eqnarray}
As noted before, the scalar fields $\chi,\rho, \eta$ may be considered as the irreducible 
components of the reducible field $\Phi(=(\chi ,\  \rho ,\ \eta )^T )$ and it is convenient to 
work in terms of hermitian scalar fields in the case of the $R_\xi$ gauges instead of 
the complex fields by writing these such as $\chi_j=\hat\chi_{2,j}+i\hat\chi_{1,j}$ with 
hermitian $\hat\chi_{1,j}$ and $\hat\chi_{2,j}$, where $j$ takes 1,2 and 3 and for instance, 
$\chi_1$ means $\chi^-$. Then it follows from App.2 that the quantities like $(\ref{37})$ 
in the scalar potential are given by the replacement of the following forms
\begin{eqnarray}
&& 
(\chi^\dag_h+\chi^\dag_v)(\chi_h+\chi_v)
\rightarrow
({\hat\chi}^{T}_h{\hat\chi}_h
+2{\hat\chi}^T_v{\hat\chi}_h
+{\hat\chi}^{T}_v{\hat\chi}_v),\nonumber\\
&&
(\chi^\dag_h+\chi^\dag_v)(\rho_h+\rho_v)
\rightarrow
({\hat\chi}^T_h+{\hat\chi}^T_v)(I_2+\tau_2)\otimes I_3
({\hat \rho}_h+{\hat \rho}_v).\label{38}
\end{eqnarray} 

Similarly, the covariant derivative for the scalar fields in $(\ref{26b})$ can be rewritten 
in terms of $\Phi$ as follows
\begin{eqnarray}
&&
D_\mu\Phi=(\partial_\mu-iK^\Phi_\mu)\Phi,\label{39}\\
&&
K^\Phi_\mu\equiv K^\chi_\mu\oplus K^\rho_\mu\oplus K^\eta_\mu,\nonumber
\end{eqnarray}
where $\oplus$ means a direct sum of the matrices and the notation $K^\varphi_\mu$ denotes 
the quantity $g\lambda\cdot A_\mu /2+g_Ny_\varphi B_\mu$ given in 
$(\ref{26b})$ with $\varphi\ (\ =\chi,\ \rho$ and $ \eta)$. $(\ref{39})$ is rewritten in 
the covariant derivative $\hat D_\mu$ corresponding to $D_\mu$ given in App.2 as follows
\begin{eqnarray}
\hat D_\mu\hat\Phi=(\partial_\mu-iK^{\hat\Phi}_\mu)\hat\Phi,\label{310}
\end{eqnarray}
where 
\begin{eqnarray*}
&&
K^{\hat\Phi}_\mu\equiv I_2\otimes(K^\chi_{\mu-} 
\oplus K^\rho_{\mu-}\oplus K^\eta_{\mu-})
+\tau_2\otimes(K^\chi_{\mu+}\oplus K^\rho_{\mu+}\oplus K^\eta_{\mu+}),\\ 
&&
K^\varphi_{\mu\pm}\equiv\frac{1}{2}(K^\varphi_\mu\pm K^{\varphi T}_\mu),
\end{eqnarray*}
and $\hat{\Phi}=(\hat{\chi}, \hat{\rho}, \hat{\eta})^T$ with $\hat{\varphi}
=(\varphi_{1,1},\varphi_{2,1},\varphi_{1,2},\varphi_{2,2},\varphi_{1,3},\varphi_{2,3})$ 
for $\varphi 
=\chi,\rho$ and $\eta$. Then, the covariant kinetic energy term 
for the scalar field $\Phi$ becomes from $(\ref{26b})$ and (B.4) in terms of $\hat\Phi$ as follows
\begin{eqnarray}
&&
({\cal D}^\mu\chi)^\dag({\cal D}_\mu\chi)
+({\cal D}^\mu \rho)^\dag ({\cal D}_\mu \rho)
+({\cal D}^\mu \eta)^\dag ({\cal D}_\mu \eta)\nonumber\\
&&={(D^\mu\Phi)}^\dag D_\mu\Phi\nonumber\\
&&=\frac{1}{2}(\hat D^\mu\hat\Phi)^T(I_2+\tau_2)
\otimes(I_3 \oplus I_3 \oplus I_3)\hat D_\mu\hat\Phi,\label{311}
\end{eqnarray}
where $I_3$ denotes $3\times 3$ unit matrix. By using the $VEV$ of the $\chi,\ \rho$ and $\eta$ 
and writing $\hat\Phi=v+\tilde\Phi$ and $v=(\hat\chi_v ,\ \hat\rho_v, \ \hat\eta_v)^T$ with 
the non-zero components $\hat\chi_{v2,3}=\chi_v,\hat\rho_{v2,2}=\rho_v, \hat\eta_{v2,1}=\eta_v$, 
$(\ref{311})$ is rewritten as follows
\begin{eqnarray}
(D^\mu\Phi)^\dag D_\mu\Phi
&& 
=\frac{1}{2}[(\hat D^\mu\tilde\Phi)^T \hat D_\mu\tilde\Phi +\{v^T(I_2+\tau_2)\otimes
(I_3\oplus I_3\oplus I_3)K^{\hat\Phi\mu}K^{\hat\Phi}_\mu)\tilde\Phi\nonumber\\
&&
+{\tilde\Phi}^T K^{\hat\Phi\mu}K^{\hat\Phi}_\mu(I_2+\tau_2)\otimes
(I_3\oplus I_3\oplus I_3)v\}-2iv^T\partial^\mu{K}^{\hat\Phi}_\mu{\tilde\Phi}\nonumber\\
&&
+v^T K^{\hat\Phi\mu}{{K}^{\hat\Phi}_\mu}v].\label{312}
\end{eqnarray}
The last term gives the mass to the gauge bosons in the form
\begin{eqnarray}
&&
[M^2_W W^{\mu +}W^-_\mu+M^2_X X^{\mu +}X^-_\mu
+M^2_Y Y^{\mu ++}Y^{--}_\mu +h.c.]\nonumber\\
&&\hspace{2cm}
+\frac{1}{2}(M^2_{Z_1}Z_1^{\mu}Z_{1\mu}
+M^2_{Z_2}Z_2^{\mu}Z_{2\mu}),\label{313}
\end{eqnarray}
where 
\begin{eqnarray*}
&&
Z_{1\mu}=\cos\phi Z_\mu+\sin\phi Z^\prime_\mu,
\hspace{0.5cm}
Z_{2\mu}=-\sin\phi Z_\mu+\cos\phi Z^\prime_\mu,\\
&&
M^2_W=\frac{g^2}{4}(\rho^2_v +\eta^2_v ),
\ \ \ 
M^2_X=\frac{g^2}{4}(\chi^2_v+\eta^2_v),\\
&&
M^2_Y=\frac{g^2}{4}(\chi^2_v+\rho^2_v),\\
&&
M^2_{Z_1}=\frac{1}{2}
\left[
M^2_Z+M^2_{Z^\prime}-\sqrt{(M^2_{Z^\prime}-M^2_Z)^2+4(M_{ZZ^\prime})^4 }
\right],\\
&&
M^2_{Z_2}=\frac{1}{2}
\left[
M^2_Z+M^2_{Z^\prime}+\sqrt{(M^2_{Z^\prime}-M^2_Z)^2+4(M_{ZZ^\prime})^4}
\right],\\
&&
\tan\phi=\frac{M^2_Z -M^2_{Z_1}}{M^2_{ZZ^\prime}},\\
&&
M^2_Z=\frac{M^2_W}{c^2_W},\\
&&
M^2_{Z^\prime}=\frac{1}{1-3t^2_W}
\left[
(2+3t^2_W)M^2_Y+(2-3t^2_W)M^2_X -M^2_W
\right],\\
&&
M^2_{ZZ^\prime}=\frac{1}{c_W\sqrt{3(1-3t^2_W)}}
\left[
M^2_Y -M^2_X +3t^2_WM^2_W
\right],
\end{eqnarray*}
$A_\mu$ denotes the massless gauge field (electromagnetic field) and $\theta_W$ the Weinberg 
angle in SM. It is noted that the neutral gauge bosons $Z_1,Z_2$ with the definite mass 
is related with $Z, Z^\prime$ through an orthogonal transformation\cite{12}.

The mass of the fermions is determined from the Yukawa interaction 
$(\ref{210})$ through the SSB. 
The lepton masses are determined from the first three terms in $(\ref{210})$ in a similar way as 
in SM and the Lagrangian is given as follows
\begin{eqnarray}
{\cal L}^l_Y
&&
=\overline{E_L}M_EE_R
+\overline{e_L}M_ee_R
+\overline{\nu_L}M_\nu E_\nu\nonumber\\
&&
+\frac{1}{\chi_v}
\left[
\overline{\nu_L}U^{\nu E}M_EE_R\chi^+_h
+\overline{e_L}U^{eE}M_EE_R\chi^{0}_h
+\overline{E_L}M_EE_R\chi^0_h
\right]\nonumber\\
&&
+\frac{1}{\rho_v}
\left[
\overline{\nu_L}U^{\nu e}M_ee_R\rho^+_h
+\overline{e_L}M_ee_R\rho^{0}_h
+\overline{E_L}U^{Ee}M_ee_R\rho^{++}_h
\right]\nonumber\\
&&
+\frac{1}{\eta_v}
\left[
\overline{\nu_L}M_\nu \nu_R\eta^0_h
+\overline{e_L}U^{e\nu}M_\nu \nu_R\eta^{-}_h
+\overline{E_L}U^{E\nu}M_\nu \nu_R\eta^+_h
\right]
+h.c.\label{314}\\
&&
=\frac{1}{\chi_v}
\left(
\overline{(U^{E\nu}\nu)},\ 
\overline{(U^{Ee}e)},\ 
\overline{E}
\right)
_LM_EE_R(\chi_v+\chi_h)\nonumber\\
&&
+\frac{1}{\rho_v}
\left(
\overline{(U^{e\nu}\nu )},\ 
\overline{e},\ 
\overline{(U^{eE}E)}\ 
\right)
_LM_ee_R(\rho_v+\rho_h)\nonumber\\
&&
+\frac{1}{\eta_v}
\left(
\overline{\nu},\ 
\overline{(U^{\nu e}e)},\ 
\overline{(U^{\nu E}E)}
\right)
_LM_\nu\nu_R(\eta_v+\eta_h)+h.c.,\nonumber
\end{eqnarray}
where 
\begin{eqnarray*}
&&
E^0_{L,R}=A^E_{L,R}E_{L,R},\ \ \  
e^0_{L,R}=A^e_{L,R}e_{L,R},\ \ \  
\nu^0_{L,R}=A^\nu_{L,R}\nu_{L,R},\\ 
&&
A^{E\dag}_{L,R}\mu^\chi A^E_{L,R}
=M_E
={\rm diag}(m_{E_1},\ m_{E_2},\ m_{E_3}),\ \ \   
E^0_{L,R}=(E^0_1,E^0_2,E^0_3)^T_{L,R},\\
&&
U^{Ee}=A^{E\dag}_LA^e_{L} 
=(U^{eE})^\dag,\ U^{\nu e}
=U^{\nu E}U^{Ee},\\
&&
\overline{E^0}_{L,R}{E^0}_{L,R}
=\overline{UE}_{L,R}{UE}_{L,R}
=\overline{E}_{L,R}{E}_{L,R}\ \ \  
{\rm for\ any\ unitary}\ U,
\end{eqnarray*}
and similar notations for the other quantities are obvious. The number of the parameters in 
$(\ref{314})$ is twenty in all and two of them are through the phases in the unitary matrices 
$U^{Ee}$ and $U^{E\nu}$\cite{13}. It follows from the nature of the mass eigenstates and 
$(\ref{314})$ that the kinetic energy parts and the Yukawa interactions for the leptons 
in terms of the mass eigenstates have the same form as those in terms of the weak interaction 
bases. The fact is due to the assignment of only the $3$ representation of $\rm SU(3)_L$ to 
the left-handed leptons and the singlet under the $3$ representation for the right-handed ones. 
It is noted that the transformation of the mass eigenstates for the leptons under 
$\rm SU(3)_L\otimes U(1)_N$ is formally expressed in the similar form as to 
the weak interaction bases, e.g.,
\begin{eqnarray}
l^\prime_{aL}\sim (1-i\sum_{b=1}^{9}\beta^bL^b)l_{aL},\label{314a}
\end{eqnarray}  
where $l_{aL}$ may have the form such as $((U_1\nu)_a,(U_2e)_a,(U_3E)_a)$ with some unitary $U$'s.

The interaction of the quarks with the $\chi$ in $(\ref{210})$ gives the mass to the $J$ quarks 
and the Lagrangian has the explicit form in terms of the masses and their eigenstates
\begin{eqnarray}
{\cal L}^\chi_Y
&&
=m_{J_3}\overline{J_{3L}}J_{3R}+\overline{J_L}M_JJ_R \nonumber\\
&&
+\frac{1}{\chi_v}m_{J_3}
\left[
\overline{u_{3L}}\chi^-_h
+\overline{d_{3L}}\chi^{--}_h
+\overline{J_{3L}}\chi^0_h
\right]
J_{3R}\nonumber\\
&&
+\frac{1}{\chi_v}
\left[
\overline{d_L}U^{dJ}M_J\chi^{-*}_h
+\overline{u_L}U^{uJ}M_J\chi^{--*}_h
+\overline{J_L}M_J\chi^{0*}_h
\right]
J_R+h.c.\label{315}\\
&& 
=\frac{1}{\chi_v}
\left(
\overline{u_3},\overline{d_3},\overline{J_3}
\right)
_Lm_{J_3}J_{3R}(\chi_v+\chi_h)\nonumber\\
&&
+\frac{1}{\chi_v}
\left(
\overline{(U^{Jd}d)},\ \overline {(U^{Ju}u)},\ \overline {J}
\right)
_LM_JJ_R(\chi_v+\chi^*_h)+h.c.,\nonumber
\end{eqnarray}
where $m_{J_3}(=\chi_v \Gamma^\chi_3 /\sqrt 2)$ denotes the mass of $J_3(=J^0_3)$, and 
the $2\times 2$ mass matrix is diagonalized in the usual way
\begin{eqnarray*}
&&
J^0_{L,R}
=A^J_{L,R}J_{L,R},
\ 
u^0_{L,R}
=A^u_{L,R}u_{L,R},
\ 
d^0_{L,R}
=A^d_{L,R}d_{L,R}, \\
&&
A^{J\dag}_LM^\chi A^J_R
=M_J\equiv
\left( 
\begin{array}{cc} 
m_{J_1} & 0 \\ 
0 & m_{J_2} 
\end{array} 
\right),\\
&& 
J^0_{L,R}=
\left( 
\begin{array}{c} 
J^0_1 \\ 
J^0_2 
\end{array} 
\right)
_{L,R},
\ 
M^\chi=\frac{\chi_v}{\sqrt 2}
\left( 
\begin{array}{cc} 
\Gamma^\chi_{11} & \Gamma^\chi_{12} \\ 
\Gamma^\chi_{21} & \Gamma^\chi_{22} 
\end{array} 
\right),\\
&&
U^{Jd}=A^{J\dag}_LA^d_L,
\ \ \ \ \ \ 
U^{Ju}=A^{J\dag}_LA^u_L,
\ \ \ \ \ \ 
A^{J\dag}_{L,R}A^J_{L,R}=I,
\ \ \ \ \ 
{\rm etc}.,\\
&&
J_{3L,R}^{0\dag}J_{3L,R}^{0}=J_{3L,R}^{\dag}J_{3L,R},
\ \ \ \ \ \ 
J_{L,R}^{0\dag}J_{L,R}^{0}=J_{L,R}^{\dag}J_{L,R}
\end{eqnarray*}
The quantities $U^{Jd}$ and $U^{Ju}$ denote the $2\times 2$ Cabibbo matrices\cite{14} and 
each one is characterized only by  one real parameter. The number of the parameters in 
$(\ref{315})$ is five and the CP violation phases do not appear\cite{13}. 
It is noted that the result is independent of the nature of the singlet because the charge of 
$J_3$ is different from that of the $J_1,\ J_2$ and thus $J_3$ can not be mixed with $J_1$ and $\ J_2$. 
The expression $(\ref{315})$ has the form similar to that in $(\ref{210})$ and 
the quantities ${(u_3,d_3,J_3)}^T_L$ and $((U^{Jd}d)_i,(U^{Ju}u)_i,J_i)^T_L$ corresponding to 
those in the weak interaction bases as well as the right-handed ones are subject to 
the transformation corresponding to that in the weak interaction bases.  

In the case (i) of $``1"$, the masses of the $u$ and $d$ quarks are given from ${\cal L}_Y$ in 
$(\ref{210})$ through the VEV of the $\rho$ and $\eta$. The Lagrangian giving the mass 
to these quarks has the form
\begin{eqnarray}
\frac{1}{\sqrt 2}
\left(
\eta_v{\bar u^0_{3L}}\Gamma^\eta_{3a}u^0_{aR}
+\rho_v{\bar d^0_{3L}}\Gamma^\rho_{3a}d^0_{aR}
+\rho_v{\bar u^0_{iL}}\Gamma^\rho_{ia}u^0_{aR}
+\eta_v{\bar d^0_{iL}}\Gamma^\eta_{ia}d^0_{aR}
\right)
+h.c.,\label{316}
\end{eqnarray}
where the sum over $i$ is taken from $1$ to $2$, and that over $a$ from $1$ to $3$. 
It is then seen that there are two possibilities of diagonalizing the mass matrices. 
One of them is to admit the mixing of the left-handed quarks belonging to 
the different representations in order to forbid unwanted processes such as 
$u(c)\leftrightarrow t$ and $d(s)\leftrightarrow b$ with the interaction with vacuum 
(or without interaction). The other one is to forbid the mixing of the left-handed quarks 
belonging to the different representations and to admit the above unwanted processes.   

The Lagrangian for the quark sector in the first one called case (i-a) can be rewritten 
in terms of the mass eigenstates as follows
\begin{eqnarray}
{\cal L}^\rho_Y+{\cal L}^\eta_Y
&&
=\overline{U_L} M_uU_R
+\overline{D_L} M_dD_R\nonumber\\
&&
+\frac{1}{\sqrt 2}\overline{U_L}B^{u\dag}_L
\left(
I_2\Gamma^\rho\rho^{0*}_h+I_1\Gamma^\eta\eta^{0}_h
\right)
B^u_RU_R\nonumber\\
&&
+\frac{1}{\sqrt 2}\overline{D_L}B^{d\dag}_L
\left(
I_1\Gamma^\rho\rho^0_h+I_2\Gamma^\eta\eta^{0*}_h
\right)
B^d_RD_R\nonumber\\
&&
+\frac{1}{\sqrt 2}\overline{U_L}B^{u\dag}_L
\left(
I_2\Gamma^\eta\eta^{-*}_h+I_1\Gamma^\rho\rho^{+}_h
\right)
B^d_RD_R\nonumber\\
&&
+\frac{1}{\sqrt 2}\overline{D_L}B^{d\dag}_L
\left(
I_2\Gamma^\rho\rho^{+*}_h+I_1\Gamma^\eta\eta^{-}_h
\right)
B^u_RU_R\nonumber\\
&&
+\frac{1}{\sqrt 2}\overline{J_{3L}}\Gamma^\rho_1B^{d}_RD_R\rho^{++}_h
+\frac{1}{\sqrt 2}\overline{J_{3L}}\Gamma^\eta_1 B^{u}_RU_R\eta^{+}_h\nonumber\\
&&
+\frac{1}{\sqrt 2}\overline{J_{L}}A^{J\dag}_L\Gamma^\rho_2B^{u}_RU_R\rho^{++*}_h
+\frac{1}{\sqrt 2}\overline{J_{L}}A^{J\dag}_L\Gamma^\eta_2 B^{d}_RD_R \eta^{+*}_h
+h.c.,\label{317}
\end{eqnarray}
where 
\begin{eqnarray*}
&&
\Gamma^{\rho,\eta}_1=
(\Gamma^{\rho,\eta}_{31}\ \Gamma^{\rho,\eta}_{32}\ \Gamma^{\rho,\eta}_{33}),
\quad
\Gamma^{\rho,\eta}_2=
\left( 
\begin{array}{ccc} 
\Gamma^{\rho,\eta}_{11} & \Gamma^{\rho,\eta}_{12} &  \Gamma^{\rho,\eta}_{13}) \\
\Gamma^{\rho,\eta}_{21} & \Gamma^{\rho,\eta}_{22} &  \Gamma^{\rho,\eta}_{23} 
\end{array} \right) , 
\quad
\Gamma^{\rho,\eta}=
\left( 
\begin{array}{c} 
\Gamma^{\rho,\eta}_{2} \\  
\Gamma^{\rho,\eta}_{1}  
\end{array} 
\right) ,\\
&& 
U^0_{L,R}=
\left( 
\begin{array}{c} 
u^0_1 \\ 
u^0_2 \\ 
u^0_3 
\end{array} 
\right)
_{L,R},
\qquad 
D^0_{L,R}=
\left( 
\begin{array}{c} 
d^0_1 \\ 
d^0_2 \\ 
d^0_3 
\end{array} 
\right)_{L,R},
U^0_{L,R}=B^u_{L,R}U_{L,R}, 
\qquad 
D^0_{L,R}=B^d_{L,R}D_{L,R},\\
&&
\frac{1}{\sqrt 2}B^{u\dag}_L
\left(
I_2\Gamma^\rho\rho_v+I_1\Gamma^\eta \eta_v
\right)
B^u_R=M_u\equiv
\left( 
\begin{array}{ccc} 
m_{u_1} & 0 & 0 \\ 
0 & m_{u_2} & 0 \\ 
0 & 0 & m_{u_3}  
\end{array} 
\right) ,\ \ \ 
I_1=
\left( 
\begin{array}{ccc} 
0 & 0 &  0 \\ 
0 & 0 & 0 \\ 
0 & 0 & 1 
\end{array} 
\right), \\
&&
\frac{1}{\sqrt 2}B^{d\dag}_L
\left(
I_1\Gamma^\rho\rho_v+I_2\Gamma^\eta\eta_v
\right)
B^d_R=M_d\equiv
\left( 
\begin{array}{ccc} 
m_{d_1} & 0 & 0 \\ 
0 & m_{d_2} & 0 \\ 
0 & 0 & m_{d_3}  
\end{array} 
\right) ,\ \ \  
I_2=
\left( 
\begin{array}{ccc} 
1 & 0 & 0 \\ 
0 & 1 & 0 \\ 
0 & 0 & 0 
\end{array} 
\right).\\  
\end{eqnarray*}
It is noted that the unitary matrices defining the mass eigenstates can not be combined into 
the Kobayashi-Maskawa-like matrices\cite{14} as in $(\ref{314})$ and $(\ref{315})$ and appear in 
the interaction by itself. It is seen that the mass matrices are diagonalized through $3\times 3$ 
unitary matrices which produce a result of the mixture of the left-handed quarks belonging to 
the different representations, $3$ and $\bar 3$, and it is necessary to diagonalize 
the $3\times 3$ matrices by $3\times 3$ unitary matrices  and use the mass eigenstates with 
three components because the quantities $\Gamma_{3i}$ and $\Gamma_{i3}(i=1,2)$ are not zero 
in general. Thus, the parameters appear through these matrices as well as those in the $\Gamma$ 
matrices and the number of the parameters in $(\ref{317})$ amounts to about fifty eight. 
The CP violation occurs through the phases in these matrices in addition to those in 
the Kobayashi-Maskawa matrices in $(\ref{314})$ for leptons\cite{13}\cite{14}. 
The following relations hold due to the unitary transformation
\begin{eqnarray}
U^{0\dag}_{L,R}U^0_{L,R}=U^{\dag}_{L,R}U_{L,R},\ \ \ \ \ \ \ \ 
D^{0\dag}_{L,R}D^0_{L,R}=D^{\dag}_{L,R}D_{L,R}.\label{318} 
\end{eqnarray}
It is evident that $(\ref{317})$ can not be rewritten in the form $(\ref{210})$ corresponding to 
that in terms of the weak interaction bases because of the mixing of the left-handed quarks 
belonging to the different representations. It is noted that the relations in $(\ref{318})$ 
do not mean the relations such as $u^{0\dag}_{3L,R} u^{0}_{3L,R}=U^{\dag}_{3L,R} U_{3L,R}$ and 
$\sum_{i=1}^2u^{0\dag}_{iL,R}u^{0}_{iL,R}=\sum_{i=1}^2U^{\dag}_{iL,R}U_{iL,R}$ 
although the reverse is true.  

The Lagrangian for the quark sector in the latter one called case (i-b) can be rewritten 
in terms of the mass eigenstates as follows
\begin{eqnarray}  
{\cal L}^\rho_Y+{\cal L}^\eta_Y
&&
=
\frac{1}{\eta_v}
\left(\overline{u_3},\overline{d_3},\overline{J_3}\right)_Lm_{u_3}u_{3R}
\left(\eta_v+\eta_h\right)
+
\frac{1}{\rho_v}
\left(\overline{u_3},\bar{d_3},\bar{J_3}\right)_Lm_{d_3}d_{3R}
\left(\rho_v+\rho_h\right)\nonumber\\
&&
+
\frac{1}{\rho_v}
\left(\overline{(U^{ud}d)},\bar{u},\overline{(U^{uJ}J)}\right)_Lm_{u}u_{R}
\left(\rho_v+\rho^*_h\right)\nonumber\\
&&
+
\frac{1}{\eta_v}
\left(\overline{d},\overline{(U^{du}u)},\overline{(U^{dJ}J)}\right)_Lm_{d}d_{R}
\left(\eta_v+\eta^*_h\right)\nonumber\\
&&
+
\frac{1}{\sqrt 2}
\left[
\left(\overline{u_3},\overline{d_3},\overline{J_3}\right)_L
\left(\Gamma^\eta_{31},\Gamma^\eta_{32}\right)A^u_Ru_R
\left(\eta_v+\eta_h\right)\right.\nonumber\\
&&
\left.
+
\left(\overline{u_3},\overline{d_3},\overline{J_3}\right)_L
\left(\Gamma^\rho_{31},\Gamma^\rho_{32}\right)A^d_Rd_R
\left(\rho_v+\rho_h\right)\right.\nonumber\\
&&
\left.
+
\left(\overline{(U^{ud}d)},\overline{u},\overline{(U^{uJ}J)}\right)_LA^{u\dag}_L
\left( 
\begin{array}{c} 
\Gamma^\rho_{13}\\ 
\Gamma^\rho_{23}
\end{array} 
\right)u_{3R}
\left(\rho_v+\rho^*_h\right)\right.\nonumber\\
&&
\left.
+
\left(\overline{d},\overline{(U^{du}u)},\overline{(U^{dJ}J)}\right)_LA^{d\dag}_L
\left( 
\begin{array}{c} 
\Gamma^\eta_{13}\\ 
\Gamma^\eta_{23}
\end{array} 
\right)d_{3R}
\left(\eta_v+\eta^*_h\right)\right]+h.c.,\label{319}
\end{eqnarray}
where   
\begin{eqnarray*} 
&&
u^0_{L,R}=A^u_{L,R}u_{L,R},
\hspace{0.5cm} 
d^0_{L,R}=A^d_{L,R}d_{L,R}, 
\hspace{0.5cm}
u^0_{3L,R}=u_{3L,R},
\hspace{0.5cm}
d^0_{3L,R}=d_{3L,R},\\
&&
u_{L,R}=
\left( 
\begin{array}{c} 
u_1 \\ 
u_2 
\end{array} 
\right)_{L,R},
\hspace{1cm} 
d_{L,R}=
\left( 
\begin{array}{c} 
d_1 \\ 
d_2 
\end{array} 
\right)_{L,R},\ \ \ \ \\
&& 
m_{u_3}=\frac{1}{\sqrt 2}\eta_v\Gamma^\eta_{33},
\ \ \ \ \ \   
m_{d_3}=\frac{1}{\sqrt 2}\rho_v\Gamma^\rho_{33},\\ 
&& 
\frac{1}{\sqrt 2}\rho_v A^{u\dag}_{L}
\left( 
\begin{array}{cc} 
\Gamma^\rho_{11} & \Gamma^\rho_{12} \\ 
\Gamma^\rho_{21} & \Gamma^\rho_{22} 
\end{array} 
\right)A^u_R=m_u=
\left( 
\begin{array}{cc} 
m_{u_1} & 0 \\ 
0 & m_{u_2}
\end{array} 
\right),\\
&&   
\frac{1}{\sqrt 2}\eta_v A^{d\dag}_{L}
\left( 
\begin{array}{cc} 
\Gamma^\eta_{11} & \Gamma^\eta_{12} \\ 
\Gamma^\eta_{21} & \Gamma^\eta_{22} 
\end{array} 
\right)A^d_R=m_d=
\left( 
\begin{array}{cc} 
m_{d_1} & 0 \\ 
0 & m_{d_2} 
\end{array} 
\right),\\
&&
U^{ud}\equiv A^{u\dag}_LA^d_L=(U^{du})^\dag,
\hspace{1cm} 
U^{Ju}\equiv A^{J\dag}_LA^u_L,
\hspace{1cm} 
U^{Jd}=U^{Ju }U^{ud},\\ 
&&
u^{0\dag}_{L,R}u^0_{L,R}=u^{\dag}_{L,R}u_{L,R}=(Uu)^{\dag}_{L,R}(Uu)_{L,R},\\
&&
d^{0\dag}_{L,R}d^0_{L,R}=d^{\dag}_{L,R}d_{L,R}=(Ud)^{\dag}_{L,R}(Ud)_{L,R}.     
\end{eqnarray*}
The Lagrangian $(\ref{319})$ has the form corresponding to that in $(\ref{210})$ in terms of 
the weak interaction bases and it is evident that the transformation of the mass eigenstate 
quarks under $SU(3)_L\otimes U(1)_N$ is formally given by the form corresponding to those in 
the weak interaction bases. It is noted that the bracketed terms on the right side 
in $(\ref{319})$ contain unwanted processes of the flavor change in the interaction with 
the vacuum.    

In the case (ii) of $``1_3"$ and $``1_{\bar 3}"$ for the singlet, however, the corresponding 
Lagrangian is given by putting $\Gamma^{\rho,\eta}_{3i}=\Gamma^{\rho,\eta}_{i3}=0\ (i=1,2)$ 
from $(\ref{319})$, i.e. given by omitting the bracketed terms in $(\ref{319})$. 
Then, it is noted that the matrices defining the mass eigenstates are combined into 
the three Cabibbo matrices one of which is given by the product of the other two and 
the Lagrangian is given in the form similar to that for the $J$ quark. Of course, the above 
unwanted processes do not appear. The number of the parameters in in this case is eight and 
no CP violation occurs. It is noted that the interaction terms are expressed 
in the forms similar to those in $(\ref{210})$ in the case (ii) also as in the case of $J$'s. 
It follows that the expression for the quark parts is summarized in terms of the mass eigenstates for 
the basic fermions and $2\times 2$ Cabibbo matrices and thus the parameters appear through 
the masses of the fermions and the Cabibbo matrices together with no $CP$ violation from the quark parts. 
Thus, the number of the parameters is twenty six in all. 
The relations $u^{0\dag}_Lu^0_L=u^{\dag}_Lu_L$ and $d^{0\dag}_Ld^0_L=d^{\dag}_Ld_L$ 
but not $(\ref{318})$ are desirable from a physical point of view as mentioned before and 
are necessary for the anomaly coefficients\cite{5} after the SSB to have the expressions 
corresponding to those in terms of the representation matrices before the SSB.
\section{$R_\xi$ gauges}\label{gau}

It is useful to use the tilde quantities for the scalar fields in order to fix the gauge and 
introduce the ghost fields\cite{9}\cite{10}. We first rewrite $(\ref{311})$ in terms of these 
quantities. Then, the non-zero components of the scalar fields in the case of the Higgs mechanism 
are given by putting ${\tilde\chi_{\alpha, i}}=0$ except for $\tilde \chi_{2,3}$ and 
${\tilde\rho}_{\alpha ,1}={\tilde\rho}_{1,2}=0$ in $(\ref{312})$ but then the terms such as 
$g\rho_v(\partial^\mu Y^{++}_\mu\rho^{++\dag}_h-\partial^\mu Y^{--}_\mu\rho^{++}_h)$ and 
$g\eta_v(\partial^\mu W^{-}_\mu\eta^{+}_h-\partial^\mu W^{+}_\mu\eta^{-}_h
+\partial^\mu X^{+}_\mu\eta^{-}_h-\partial^\mu X^{-}_\mu \eta^{+}_h)$ , which lead to 
the unwanted processes such as $\rho^{--}\leftrightarrow Y^{--}$ 
and $\eta^{+}\leftrightarrow W^{+}$ for moving gauge bosons, remain in the third term in 
$(\ref{312})$. These terms can not be excluded with some additional physical procedure as long as 
the Higgs mechanism is used. Therefore, it is more reasonable to use $R_\xi$ gauges than 
the gauge based on the Higgs mechanism because the terms giving the above unwanted processes 
can be cancelled by terms added to the Lagrangian to fix the gauge. 

The gauge fixing Lagrangian is introduced according to the known procedure\cite{9} as follows
\begin{eqnarray}
{\cal L}_{gf}=-\frac{\xi}{2}\sum_{a=1}^9
\left(
\partial^\mu V^a_{\mu}-i\frac{g^a}{\xi}v^T{L^{\hat\Phi a}}{\tilde\Phi}
\right)^2\ ,\label{41}
\end{eqnarray}
where 
\begin{eqnarray*}
&&
L^{\hat\Phi a}\equiv
 I_2 \otimes(L^{\chi a}_-\oplus L^{\rho a}_-\oplus L^{\eta a}_-)
+\tau_2 \otimes(L^{\chi a}_+\oplus L^{\rho a}_+\oplus L^{\eta a}_+),\\
&&
L^{\varphi a}_{\pm}\equiv\frac{1}{2}
\left(L^{\varphi a}\pm L^{\varphi aT}\right),\\
&&
L^{\varphi a}=L^a=
\left\{
\begin{array}{cc}
\lambda^a/2 & (a=1,2,\cdots,8),\\ 
y_\varphi I_3 & (a=9),
\end{array} \right.                  
\end{eqnarray*}
and $V^a_\mu =A^a_\mu ,\ g_a =g$ ($a =1,2,\cdots, 8$),\ $V^9_\mu=B_\mu$ and $g_9=g_N$. 
It is noted that the cross terms from $(\ref{41})$ cancel the third term in $(\ref{312})$. 
Furthermore, as is well known a set of nine ghost fields $C^a (a=1,2,\cdots , 9)$ 
corresponding to the nine generators $L_i=\lambda_i/2 (i=1,2,\cdots ,8)$ and $ y_\phi I_3$ 
are needed to ensure unitarity and renormalizability\cite{9}\cite{10}\cite{15}. 
The Lagrangian for the ghost fields is given with the known procedure\cite{9} by
\begin{eqnarray}
{\cal L}_{gh}=i\left[
\partial^\mu\overline{C^a}(\partial_\mu C^a + g_bf_{abc}V_\mu^bC^c )
-\frac{ g_a  g_b}{\xi}\overline{C^a}v^TL^{\hat\Phi a}L^{\hat \Phi b}
(v+{\tilde \Phi})C^b \right],\label{42}
\end{eqnarray}
where $f_{abc}$ denotes the structure constants of $SU(3)_L\otimes U(1)$ with $f_{ab9}=0$. 
The mass term of the ghost fields contained in the last term of $(\ref{42})$ is given by 
replacing $V^a_\mu \rightarrow C^a$ in that of the gauge fields given by the last term in 
$(\ref{312})$ except for the factor $\xi$. The explicit expression of $(\ref{42})$ in terms of 
the $\chi ,\ \rho,$ and $\eta $ 
is given as follows
\begin{eqnarray}
{\cal L}_{gh}
&&
=i\left[
\partial^\mu\overline{C^j}(\partial_\mu C^j+gf_{jkl}A_\mu^kC^l)
+\partial^\mu\overline{C}\partial_\mu C\right.\nonumber\\
&&
\left.
-\frac{g^2}{2\xi}\overline{C^j}
\left\{\chi_v^T
\left\{
I_2\otimes(L^jL^k+L^{jT}L^{kT})
+\tau_2\otimes(L^jL^k-L^{jT}L^{kT})
\right\}(\chi_v+{\tilde\chi})\right.\right.\nonumber\\
&&
\left.\left.
+\rho_v^T\left\{
I_2 \otimes(L^jL^k+L^{jT}L^{kT})
+\tau_2\otimes(L^jL^k-L^{jT}L^{kT})\right\}
(\rho_v+{\tilde\rho})\right.\right.\nonumber\\
&&
\left.\left.
+\eta_v^T\left\{
I_2 \otimes(L^jL^k+L^{jT}L^{kT})
+\tau_2\otimes(L^jL^k-L^{jT}L^{kT})\right\}
(\eta_v+{\tilde\eta})\right\}C^k\right.\nonumber\\
&&
\left.
-\frac{g_N^2}{\xi}\overline{C}\left\{
\chi_v^T(\chi_v+\tilde\chi)
+\rho_v^T(\rho_v+\tilde\rho )\right\}C\right.\nonumber\\
&&
\left.
+\frac{gg_N}{2\xi}\overline{C^j}\left\{
\chi_v^T\left\{
\tau_2\otimes(L^j-L^{jT})+I_2\otimes(L^j+L^{jT})\right\}
(\chi_v+{\tilde\chi})\right.\right.\nonumber\\
&&
\left.\left.
+\rho_v^T\left\{
\tau_2\otimes(L^j-L^{jT})
+I_2\otimes(L^{j}+L^{jT})\right\}
(\rho_v+{\tilde\rho})\right\}C\right.\nonumber\\
&&
\left.
+\frac{gg_N}{2\xi}\overline{C}\left\{
\chi_v^T\left\{\tau_2\otimes(L^j-L^{jT})
+I_2\otimes(L^{j}+L^{jT})\right\}(\chi_v +{\tilde\chi})
\right.\right.\nonumber\\
&&
\left.\left.
+\rho_v^T\left\{
\tau_2\otimes(L^j-L^{jT})+I_2\otimes(L^{j}+L^{jT})\right\}
(\rho_v+{\tilde\rho})\right\}C^j\right] ,\label{43}
\end{eqnarray}
where $C \equiv C^9$ and $\chi_v (\rho_v, \eta_v)$ is used instead of 
$\hat {\chi}_v \ (\hat {\rho}_v,\ \hat {\eta}_v)$. 
\section{Propagators}\label{pro}

As the Lagrangian in the $R_\xi$ gauge is determined, the propagators for the particles are 
obtained in a known way. Those for the gauge bosons are obtained in the form similar to those 
in SM from ${\cal L}_V$ in $(\ref{26c})$ and $(\ref{312})$ as follows
\begin{eqnarray}
&&
iD^A_{\mu\nu}(k)=-i\frac{1}{k^2}\left[
g_{\mu\nu}-\frac{(1-\frac{1}{\xi})}{k^2}k_\mu k_\nu
\right], \hspace{2.3cm}{\rm for}\ A_\mu\nonumber\\
&&
iD^V_{\mu\nu}(k)=-i\frac{1}{k^2-M^2_V}\left[
g_{\mu\nu}-\frac{(1-\frac{1}{\xi})}{k^2-M^2_V/\xi}k_\mu k_\nu
\right], \quad {\rm for}\ 
V=W^{\pm}_\mu, X^{\pm}_\mu,Y^{\pm \pm}_\mu,\nonumber\\
&&
iD^{Z_i}_{\mu\nu}(k)=-i\frac{1}{(k^2-M^2_{Z_i})}\left[
g_{\mu\nu}-\frac{(1-\frac{1}{\xi})k_\mu k_\nu}{(k^2-M^2_{Z_i}/\xi)}
\right],\quad {\rm for}\ Z_i\ (i=1,2),\label{51}
\end{eqnarray}
where here and hereafter a factor to select the boundary condition is omitted.

Those for the ghost fields are given from those in $(\ref{43})$ which has the same mass matrix as 
the gauge field in $(\ref{313})$ except for the parameter $\xi$ and thus the states of 
the ghost may be classified in the way similar to those of the gauge bosons. 
The notations corresponding to the gauge bosons are used,i.e., such as $c^+_W,c^+_X,c_A ,etc.$ 
corresponding to $W^+_\mu,X^+_\mu, A_\mu ,etc.$ and the propagators are given as follows
\begin{eqnarray}
&&
iD^G_A(k)=\frac{i}{k^2}, \hspace{2.5cm} {\rm for}\ c_A\nonumber\\
&&
iD^G_X(k)=\frac{i}{k^2-M^2_X/\xi}, \qquad {\rm for}\ 
X=c^\pm_W, c^\pm_X,c^{\pm \pm}_Y, \label{52}\\
&&
iD^G_{Z_i}(k)=\frac{i}{k^2 -M^2_{Z_i}/\xi}, \qquad {\rm for}\ 
c_{Z_i}\ (i=1,2),\nonumber
\end{eqnarray}
where the expressions for $c^+_W$ etc. in terms of $c^i$ etc. are given by replacing 
the $A^i_\mu$ and $B_\mu$ in the expressions for the gauge bosons $W^+_\mu$ etc. with $c^i$ 
and $c$ and the $M^2_X$ denotes the mass of the gauge boson $X$. It is noted that the first one 
in $(\ref{52})$ corresponds to a propagator of the mass zero field and remains in the limit of 
$\xi \rightarrow 0$ in contrast to those of the other eight.

The propagators for the scalar fields are divided into three types and are given as follows:

\noindent
(I)Type 1. The mass matrix for a pair of fields 
$(\tilde x, \tilde y)$ is given by the $2\times 2$ matrix
\begin{eqnarray*}
\mu^2+\frac{M^2}{4\xi}=c
\left( 
\begin{array}{c} 
y \\ 
-x 
\end{array}
\right)(y\ -x )
+\frac{g^2}{4\xi}\left( 
\begin{array}{c} 
x \\ 
y 
\end{array}
\right)(x\ y ).
\end{eqnarray*}
The propagators for the diagonalized fields of these are given by

\noindent
(a) For $\frac {g^2}{4\xi}>c$;
\begin{eqnarray}
&&
iD^{\hat x}_{sc}(k)=\frac{i}{k^2-\frac{4cM^2_x}{g^2}},
\hspace{1cm} 
iD^{\hat y}_{sc}(k)=\frac{i}{k^2-\frac{M^2_x}{\xi}},\label{53}\\
&&
\hat x=\cos\phi_s\tilde x-\sin\phi_s\tilde y ,
\hspace{1cm} 
\hat y=\cos\phi_s\tilde y+\sin\phi_s\tilde x ,\nonumber\\
&&
M^2_x=\frac{g^2}{4}(x^2+y^2),
\hspace{1cm} 
\tan\phi_s=\frac{y}{x},\nonumber
\end{eqnarray}
(b) For $c>\frac {g^2}{4\xi}$;

\noindent
The propagator in this case is given by replacing $x\rightarrow y$ and $y\rightarrow -x$ in (a).

It follows that one of the two propagators in $(\ref{53})$ has a physical pole $4cM^2_x/g^2$ 
depending on a mass $M^2_x$ of the gauge boson $x$ and the other a pole $M^2/\xi$ depending on 
the $\xi$. Thus one of these remains even in a limit $\xi \rightarrow 0$ and the other 
disappears in the limit. Explicitly, the following six cases are possible for a pair of fields
\begin{eqnarray*}
&&
({\rm i})\hspace{0.5cm}x=\chi _v,  y = \eta_v, c=c_8/2, M_x^2 =M^2_X\hspace{0.8cm}for\ \tilde {x} 
=\tilde {\chi }_{1,1},\tilde {y} =\tilde {\eta}_{1,3}.\\
&&
({\rm ii})\hspace{0.4cm}x=\eta _v,  y =- \chi_v, c=c_8/2, M_x^2 =M^2_X\hspace{0.5cm}for\ \tilde {x} 
=\tilde {\chi }_{2,1},\tilde {y} =\tilde {\eta}_{2,3}.\\
&&
({\rm iii})\ \ x=\chi _v,  y = \rho_v, c=c_7/2, M_x^2 =M^2_Y\hspace{0.8cm}for\ \tilde {x} 
=\tilde {\chi }_{1,2},\tilde {y} =\tilde {\rho}_{1,3}.\\
&&
({\rm iv})\ \ x=\chi _v,  y = -\rho_v, c=c_7/2, M_x^2 =M^2_Y\hspace{0.5cm}for\ \tilde {x} 
=\tilde {\chi }_{2,2},\tilde {y} =\tilde {\rho}_{2,3}.\\
&&
({\rm v})\hspace{0.4cm}x=\rho _v,  y = \eta_v, c=c_9/2, M_x^2 =M^2_W\hspace{0.8cm}for\ \tilde {x} 
=\tilde {\rho }_{1,1},\tilde {y} =\tilde {\eta}_{1,2}.\\
&&
({\rm vi})\ \ x=\rho _v,  y =- \eta_v, c=c_9/2, M_x^2 =M^2_W\hspace{0.5cm}for\ \tilde {x} 
=\tilde {\rho }_{2,1},\tilde {y} =\tilde {\eta}_{2,2}.
\end{eqnarray*}
\noindent
(II)Type 2. The mass matrix for the 
$( \tilde {\chi}_{1,3},\tilde {\rho}_{1,2},\tilde {\eta}_{1,1})$ is given by
\begin{eqnarray}
M^2=\frac{1}{\xi}
\left( 
\begin{array}{ccc} 
(\frac{g^2}{3}+g^{\prime 2})\chi^2 & 
-(\frac{g^2}{6}+g^{\prime 2})\chi \rho & 
-\frac{g^2}{6}\chi \eta \\  
-(\frac{g^2}{6}+g^{\prime 2})\chi \rho & 
(\frac{g^2}{3}+g^{\prime 2})\rho^2 & 
-\frac{g^2}{6}\rho\eta \\ 
-\frac {g^2}{6}\chi \eta & 
-\frac {g^2}{6}\rho \eta &  
\frac {g^2}{3}\eta^2 
\end{array}\right),\label{54}
\end{eqnarray}
where the notations $(\chi , \rho, \eta )$ are used instead of $(\chi_v , \rho_v, \eta_v)$. 
The propagators are given for the diagonalized fields of these by
\begin{eqnarray*}
&&
iD_{0}(k)=\frac{i}{k^2},\ \ \ \ \ \ 
iD_{-}(k)=\frac{i}{k^2-M^2_{-}/\xi},\ \ \ \ \ \ 
iD_{+}(k)=\frac{i}{k^2-M^2_{+}/\xi},
\end{eqnarray*}
where 
\begin{eqnarray*}
&&
{\tilde \chi}_{0}=
a(\rho \eta {\tilde \chi}_{1,3} 
+\chi \eta {\tilde \rho}_{1,2}
+\chi \rho {\tilde \eta}_{1,1}),\ \ \ \ \ 
{\tilde \rho}_{M_+}=
\cos \theta {\tilde \rho}_{-}
+\sin \theta {\tilde \eta}_{+},\ \ \ \ \\
&&
{\tilde \eta}_{M_-}=
-\sin \theta  {\tilde \rho}_{+}
+\cos \theta{\tilde \eta}_{-},\ \ \ \ \ 
{\tilde \rho}_{-}=
b(\chi{\tilde \chi}_{1,3} 
+\rho {\tilde \rho}_{1,2}
-2\eta {\tilde \eta}_{1,1}),\ \ \ \ \\
&& 
{\tilde \eta}_{+}=
c\{\chi (\rho^2+2\eta^2){\tilde\chi}_{1,3} 
-\rho (\chi^2+2\eta^2){\tilde \rho}_{1,2}
+\eta (\chi^2-\rho^2){\tilde \eta}_{1,1}\},\\
&&
a^2(\chi^2 \rho^2+\chi^2 \eta^2+\eta^2 \rho^2)=1,\ 
b^2(\chi^2 +\rho^2 +4\eta^2)=1,\ 
c=ab,\\
&& 
\tan\theta=
\frac{1}{M^2_{23}}
\left[
M^2_{33}-M^2_{22}-\sqrt{(M^2_{33}-M^2_{22})^2+4(M^2_{23})^2}
\right],\\
&&
M^2_{22}=
b^2\left[
\frac{g^2}{3}(\chi^4+\rho^4+4\eta^4-\rho^2 \chi^2 
+2\eta^2 \chi^2+2\rho^2 \eta^2)
+g^2_N(\chi^2-\rho^2)^2
\right],\\
&&
M^2_{33}=
\frac{4b^2}{a^2}\left(\frac{g^2}{4}+g^2_N\right),\ \ \ \ \ \ \ 
M^2_{23}=
\frac{2b^2}{a}\left(\chi^2-\rho^2\right)
\left(\frac{g^2}{4}+g^2_N\right),\\
&&
M^2_\pm=\frac{1}{2}
\left[
\frac{g^2}{3}
\left(\rho^2+\chi^2+\eta^2\right)\right.\\
&&
\left.
\mp\sqrt{
\left\{\frac{g^2}{3}\left(\rho^2+\chi^2+\eta^2\right)
+g^2_N\left(\rho^2+\chi^2\right)\right\}^2
-\frac{4g^2}{3}\left(\frac{g^2}{4}+g^2_N\right)
\left(\chi^2 \rho^2+\chi^2\eta^2+\rho^2\eta^2\right)}\ \right].
\end{eqnarray*}
It is noted that one of the propagators appears with the zero-mass form as above due to 
${\rm det} M^2 =0$.

\noindent 
(III)Type 3. The propagator of the case is for the fields 
$({\tilde \chi}_{2,3}, {\tilde \rho}_{2,2},{\tilde \eta}_{2,1})$, which correspond to 
the neutral fields appearing with the masses in the Higgs mechanism and are the real counterparts 
of the neutral scalar fields whose imagianry parts appear in the type 2, and is given by
\begin{eqnarray}
iD(\tilde{\chi}_{2,3},\tilde{\rho}_{2,2},\tilde {\eta}_{2,1})
=\frac{i}{k^2 -M^2},\label{55}
\end{eqnarray}
where the mass matrix is given by
\begin{eqnarray*}
M^2=
\left( 
\begin{array}{ccc} 
2c_1\chi^2 & c_4 \chi \rho &  c_5\chi \eta \\ 
c_4\chi \rho & 2c_2 \rho^2 &  c_6\rho \eta \\ 
c_5\chi \eta & c_6 \rho \eta & 2c_3\eta^2 
\end{array} 
\right).
\end{eqnarray*}

It is noted that the first type of the propagators has two types of the pole, one at 
$4cM^2_x/g^2$ and the gauge-dependent pole at $M^2_x/\xi$ with the gauge boson mass and 
the second one has the mass matrix proportional to $\xi^{-1}$ whose determinant is zero. 
The third type is for the fields $(\chi_{2,2},\ \rho_{2,2},\ \eta_{2,1})$ and the mass matrix 
with three non-zero eigenvalues is independent of $\xi$.

It follows that in the limit of $\xi \rightarrow 0$, there are one massless (photon), and 
eight massive vector bosons, ten scalar bosons (18 - 8=10) one of which is massless 
and no other particles independent of $\xi$. The result for the degree of freedom agrees with 
that expected from the Higgs mechanism. It is, however, noted that as mentioned before 
an unwanted processes appear in the Higgs mechanism but in the $R_\xi$ gauge 
the unwanted processes are cancelled out by the gauge fixing terms. It is confirmed that all 
the above propagators behave like $( momentum )^{-2}$ in the momentum $\rightarrow \infty $ limit in 
the case of $0<\xi < \infty $. Thus it is expected that the theory with the $R_\xi$ gauge will be 
renormalizable provided that the triangle anomalies do not appear from the interaction of 
the fermion fields with those of the gauge bosons as is well known. However, it is shown without 
the gauge fixing in a previous paper\cite{6} that the nature of the singlet assigned to 
the right-handed quarks in the families affects the triangle anomalies.

The propagator for the mass eigenstate fermions can be discussed in the usual way from 
the second order terms in $(\ref{314})$ for the leptons and the discussions after $(\ref{315})$ 
for the quarks together with the kenetic energy terms expressed in terms of the mass eigenstates 
from $(\ref{26a})$. The propagators for the leptons and the $J$ quarks are given in a unique way. 
However, the propagators for the mass eigenstate $u$ and $d$ quarks have the different forms 
depending on the nature of the right-handed singlet.

In the case (i-a), the progpagator for each quark of $U_a$ and $D_a$ is given throught 
$(\ref{317})$ and the kinetic energy terms expressed in terms of the mass eigenstates from 
$(\ref{26a})$ in the same form as those for the leptons. It is, however, noted that 
the mass eigenstates in $(\ref{317})$ are given by a linear combination of the quarks belonging to 
different represenattions as mentioned below $(\ref{317})$. It,thus, follows that in this case 
the quark currents in terms of the mass eigenstates have the form different from those in terms of 
the weak interaction bases. In the case (i-b), owing to the braketed terms in $(\ref{319})$ 
the propagators are not given for each quark but with the form related to the three quarks 
such as of the $u_1,\ u_2$ and $u_3$ each of which is a linear combination of 
three weak interaction bases. The propagators for the $u$ and $d$ are obtained from 
$(\ref{319})$ and the kinetic energy terms in $(\ref{26a})$ and for instance, those for 
the $u_1$,$u_2$ and $u_3$ are given in the coupled form in the known way as follows
\begin{eqnarray}
&&
iD(k)=i
\left( 
\begin{array}{ccc} 
\ooalign{\hfil/\hfil\crcr$k$} -m_1 & 0 & h_{13} \\  
0 & \ooalign{\hfil/\hfil\crcr$k$} -m_2 & h_{23} \\ 
h_{31} & h_{32} & \ooalign{\hfil/\hfil\crcr$k$} -m_3 
\end{array} 
\right)^{-1},\nonumber\\
&&
h_{i3}\equiv\frac{1}{\sqrt 2}
\left\{
\eta_v\sum\Gamma^{\eta *}_{3j}(A^u_R)^*_{ji}P_L
+\rho_v\sum(A^u_R)^*_{ji}\Gamma^{\eta }_{j3}P_R
\right\},\nonumber\\
&&
h_{3i}\equiv\frac{1}{\sqrt 2}
\left\{
\eta_v\sum \Gamma^{\eta}_{3j}(A^u_R)_{ji}P_R
+\rho_v\sum\Gamma^{\rho *}_{j3}(A^u_L)_{ji}P_L 
\right\},\nonumber\\
&&
m_1=m_{u_1},\ m_2=m_{u_12},\ m_3=m_{u_3} ,\label{56} 
\end{eqnarray}
where the sum over j is meant from 1 to 2. It, thus, follows that the quark currents in terms of 
the mass eigenstate quarks have the same form as those in the weak interaction bases but 
the anomaly coefficients from the quark currents can not be expressed in the form with 
the product of the representation matrices in terms of the weak interaction bases (see \S\ref{tri}) 
because of the propagator by the mixing of the quarks in the different representations.

In the case (ii) of taking into account the transformation property of the right-handed singlet, i.e., 
$``1_3"$ for the counterpart of the left-handed $``3"$ representation and $``1_{\bar 3}"$ 
for that of the left-handed $``\bar 3"$ representation, the propagators are given for each quarks 
such as for the $J$ quarks and then such phenomena appearing in the case (i-b) for 
the anomaly coefficient does not exist for lack of the bracketed terms in $(\ref{319})$, i.e., 
due to $\Gamma^{\rho, \eta}_{i3}=\Gamma^{\rho, \eta}_{3i}=0$. The result will play an important role 
in the fermion triangle anomalies and thus in the renormalization of the theory.

As is seen from above discussion on the quark propagators the quantities 
$\Gamma^{\rho ,\eta }_{3i}$ and $\Gamma^{\rho ,\eta }_{i3}$ for $i=1,2$ must be zero 
in order to give the consistent result to the theory, i.e. anomaly free and then renorlizability. 
The condition of  $\Gamma^{\rho, \eta}_{i3}=\Gamma^{\rho, \eta}_{3i}=0$ is satisfied automatically 
in the case (ii) of our assignment of $``1_{3}"$ and $``1_{\bar 3}"$, but is not ensured in the case (i) of $``1"$ 
except for imposition of the condition by hand. In the next section, the fermion currents are given 
in terms of the mass eigenstates. 
\section{Fermion currents in terms of mass eigenstates}\label{fer} 

The currents in $(\ref{211})$ in terms of the weak interaction bases can be given in terms of 
the mass eigenstates given in $(\ref{314})$ for the lepton and those after $(\ref{315})$ for the quarks. 
Owing to the last relations below $(\ref{314})$ and the corresponding ones such as the relation $(\ref{318})$ 
and the last relation below $(\ref{315})$ and $(\ref{319})$ for the quarks, the lepton parts in the currents 
$J_{W}^{\mu}$ etc.,\cite{16} and the electromagnetic current $J_{em}^{\mu}$ may be given by substitution of 
the weak interaction bases in $(\ref{211})$ into the corresponding mass field or the mass field multiplied with 
some unitary matrix such as $\nu^\prime =U^{e\nu }\nu$ and $D^{\prime}=BD$ as seen from 
$(\ref{314})\sim (\ref{319})$. However, the quark parts of the currents except for the electromagnetic current 
have the different forms after the SSB according to the nature of the singlet\cite{6}. 

In the case (i-a), the quark part of the $J^\mu_W $ current can be rewritten in the form
\begin{eqnarray}
J^\mu_W
&&
\sim\overline{{Q^0}_{3L}}\gamma^\mu(\lambda_1+i\lambda_2)Q^0_{3L}
+\overline{Q^0_{jL}}\gamma^\mu(-\lambda^T_1-i\lambda^T_2)Q^0_{jL}\nonumber\\
&&
=2\overline{U_{aL}}(B^{u\dag})_{a3}\gamma^\mu({B^d_L})_{3b}D_{bL}
-2\overline{U_{aL}}(B^{u\dag})_{aj}\gamma^\mu({B^d_L})_{jb}D_{bL},\label{61}
\end{eqnarray}
where the sum over $a$ and $b$ is meant from 1 to 3 and that over $j$ from 1 to 2. 
It is noted that the last expression can not be rewritten owing to the different sign in front of 
the terms in the same form as that in the weak interaction bases. This different sign comes from the fact 
that the left-handed quarks in the first and the second families are transformed under 
the $\bar 3$ representation but those in the third family are under the $3$ representation. 
The mixing of the quarks in the different representations as mentioned below $(\ref{317})$ 
must be used to construct the mass eigenstates. Similarly, it is apparent that the quark parts 
for the other currents can not be expressed in the form corresponding to those in terms of 
the weak interaction bases and in particular the FCNC\cite{6}\cite{8} appears from the neutral current 
$J^\mu_{Z^\prime}$ such as
\begin{eqnarray*}
\overline{{Q^0}_{3L}}\gamma^\mu \lambda_8Q^0_{3L}
-\overline{{Q^0}_{jL}}\gamma^\mu \lambda^T_8Q^0_{jL}
&&
=\overline {U_{aL}}{(B^{u\dag}_L)}_{a3}\gamma^\mu{(B^u_L)}_{3b}U_{bL}
-\overline {U_{aL}}{(B^{u\dag}_L)}_{aj}\gamma^\mu{(B^u_L)}_{jb}U_{bL}\nonumber\\
&&
+\overline {D_{aL}}{(B^{d\dag}_L)}_{a3}\gamma^\mu{(B^d_L)}_{3b}D_{bL}
-\overline {D_{aL}}{(B^{d\dag}_L)}_{aj}\gamma^\mu{(B^d_L)}_{jb}D_{bL}\nonumber\\
&&
+\overline {J_{3L}}\gamma^\mu J_{3L}-\overline {J_{L}}\gamma^\mu J_{L}.
\end{eqnarray*}
The FCNC does not occur in the $J$ quarks but occurs in the $u$ and $d$ quarks 
bucause the mixing of the $J$ quarks belonging to the different representations does not appear 
as seen from $(\ref{315})$ but that of the $u$ and $d$ quarks appears as seen from $(\ref{317})$
\cite{8}\cite{12}\cite{13}. 
 
In the case (i-b), the quark part of the $J^\mu_W$ current can be given in 
the form similar to that in the weak interaction bases,i.e.,
\begin{eqnarray}
J^\mu_W
&&
\sim\overline{Q^0_{3L}}\gamma^\mu(\lambda_1+i\lambda_2)Q^0_{3L}
+\overline{Q^0_{jL}}\gamma^\mu(-\lambda^T_1-i\lambda^T_2)Q^0_{jL}\nonumber\\
&&
=\overline{Q_{3L}}\gamma^\mu(\lambda_1+i\lambda_2)Q_{3L}
+\overline{Q_{jL}}\gamma^\mu(-\lambda^T_1-i\lambda^T_2)Q_{jL},\label{62}
\end{eqnarray}
where 
\begin{eqnarray*}
&&
Q_{3L}=\left(
\begin{array}{c} 
u_3 \\  
d_3 \\ 
J_3 
\end{array} \right)_L,
\hspace{1cm} 
Q_{iL}=\left( 
\begin{array}{c} 
d_i \\  
u^\prime_{i} \\ 
J^\prime_i 
\end{array} \right)_L,\\
&&
u^\prime_L=U^{du}u_L,\hspace{1.5cm} J^\prime_L =U^{dJ}J_L.
\end{eqnarray*}
It is apparent that the other currents for the quark parts have the forms corresponding to those 
in the weak interaction bases and then the cross terms of the left-handed quarks belonging to 
the different representations do not appear in contrast to the case (i-a). In this case, however, 
the propagators  depend on the quarks belonging to the different representations as seen from 
$(\ref{56})$ in \S\ref{pro}.
  
In the case (ii) of $``1_3"$ and $``1_{\bar 3}"$, the currents in $(\ref{211})$ are given in 
the same form as in the case (i-b). A decisive difference from the case (i-b) is that in the case (i-b) 
the bracketed terms in $(\ref{319})$ appear but in the case (ii) these terms vanish due to 
$\Gamma^{\rho, \eta}_{3i}=\Gamma^{\rho, \eta}_{i3}=0$. As the result, the propagators of 
the $u$ and $d$ quarks are given for each mass eigenstate quark in contrast to those 
in the cases (i-a) and (i-b) and then the anomaly coefficients after the SSB are given in 
the same form as that in the weak interaction bases as will be seen in the next section. 
\section{Triangle anomalies}\label{tri}

It is known that the anomalies, which are singuralities associated with the fermion triangle 
diagram contributions to the vertex of three currents, should be cancelled even when 
the gauge structure of the theory is modified by the SSB\cite{5}. If the anomalies appear 
without vanishing, the renormalizability of the theory will be destroyed\cite{11}\cite{15}.

The anomaly coefficient $A_{ijk}$ in the vertex of currents $i,j$ and $k$ is given by\cite{5}
\begin{eqnarray}
A_{ijk}&&
\sim 2{\rm Tr}L^i_L\{L^j_L,\ L^k_L\}
-2{\rm Tr}L^i_R\{L^j_R,\ L^k_R\}\nonumber\\
&&\equiv 2 (A^L_{ijk}-A^R_{ijk}),\label{71}
\end{eqnarray}
where $L^i_L$ and $L^i_R$ denote the represenation matrices of 
the left- and right-handed fermions. 
In the case (ii) of the singlet $1_3$ and $1_{\bar 3}$, it follows from the discussion 
in \S\ref{fer} that the anomaly coefficients are zero because the currents after the SSB are 
expressed in terms of ones corresponding to those in terms of the weak interaction bases 
before the SSB. For instance, the coefficients for the processes $Z \rightarrow W^+W^-$ and 
$Z \rightarrow Y^{++}Y^{--}$ become
\begin{eqnarray}
&&
\sum {\rm Tr}(T_{3L}-2s^2_W Q_L)\{T_{1L}-iT_{2L},T_{1L}+iT_{2L}\}\nonumber\\
&&\hspace*{2cm}
=\frac{3}{8}[{\rm Tr}\lambda_3
\{\lambda_1-i\lambda_2,\ \lambda_1+i\lambda_2 \}\nonumber\\
&&\hspace*{2cm}
+{\rm Tr}\lambda_3\{\lambda_1-i\lambda_2,\ \lambda_1+i\lambda_2\}
+2{\rm Tr}(-\lambda^T_3)
\{-\lambda^T_1+i\lambda^T_2,\ -\lambda^T_1-i\lambda^T_2\}]\nonumber\\
&&\hspace*{2cm}
-6s^2_W[{\rm Tr}Q_l \{\lambda_1-i\lambda_2,\ \lambda_1+i\lambda_2 \}
+{\rm Tr}Q_{3L}\{\lambda_1-i\lambda_2,\ \lambda_1+i\lambda_2\}\nonumber\\
&&\hspace*{2cm}
+2{\rm Tr}(Q_{iL})
\{-\lambda^T_1+i\lambda^T_2,\ -\lambda^T_1-i\lambda^T_2\}]\nonumber\\
&&\hspace*{2cm}
=0,\nonumber\\
&&
\sum {\rm Tr}(T_{3L}-2s^2_W Q_L)\{T_{6L}-iT_{7L},T_{6L}+iT_{7L}\}\nonumber\\
&&\hspace*{2cm}
=\frac{3}{8}[{\rm Tr}\lambda_3
\{\lambda_6-i\lambda_7,\ \lambda_6+i\lambda_7 \}\nonumber\\
&&\hspace*{2cm}
+{\rm Tr}\lambda_3\{\lambda_6-i\lambda_7,\ \lambda_6+i\lambda_7\}
+2{\rm Tr}(-\lambda^T_3)
\{-\lambda^T_6+i\lambda^T_7 ,\ -\lambda^T_6-i\lambda^T_7\}]\nonumber\\
&&\hspace*{2cm}
-6s^2_W[{\rm Tr}Q_l \{\lambda_1-i\lambda_2,\ \lambda_1+i\lambda_2 \}
+{\rm Tr}Q_{3L}\{\lambda_1-i\lambda_2,\ \lambda_1+i\lambda_2\}\nonumber\\
&&\hspace*{2cm}
+2{\rm Tr}(Q_{iL})
\{-\lambda^T_1+i\lambda^T_2,\ -\lambda^T_1-i\lambda^T_2\}] \nonumber\\
&&\hspace*{2cm}
=0,\label{72}
\end{eqnarray}
where the quantities $Q_l,\ Q_{3L}$ and $Q_{iL}$ are the charges of 
the leptons  and quarks given by
\begin{eqnarray*}
Q_{l}=\left( 
\begin{array}{ccc} 
0 & 0 & 0 \\ 
0 & -1 & 0 \\ 
0 & 0 & 1 
\end{array} 
\right ),\hspace{0.5cm}
Q_{3L}=\frac{1}{3}\left( 
\begin{array}{ccc} 
2 & 0 & 0 \\ 
0 & -1 & 0 \\ 
0 & 0 & 5 
\end{array} 
\right ),\hspace{0.5cm}
Q_{iL}=\frac{1}{3}\left( 
\begin{array}{ccc} 
-1 & 0 & 0 \\ 
0 & 2 & 0 \\ 
0 & 0 & -4 
\end{array} 
\right ).
\end{eqnarray*}
and the sum is meant over the families. It is noted that the factor $3$ in $(\ref{72})$ comes 
from the three families for the leptons and the three colors for the quarks. It follows that 
the anomaly coefficients are not necessarily zero due to the sum of all charge in three families 
in contrast with the case in the SM. In this way it is easy to see that the anomaly coefficients 
vanish for all possible processes. Thus, the renormalizability of the theory will be guaranteed 
in the case (ii), while it is apparent that in the case (i) the coefficients will not become zero 
in general because the anomaly coefficients can not be expressed in the form of $(\ref{71})$ 
for both (i-a) and (i-b) due to the mixing of the quarks belonging to the different representations 
as seen from $(\ref{317})$ and $(\ref{56})$.
\section{BRS transformation}\label{brs}

In this section, we discuss a BRS transformation\cite{11}. It is well known that the invariance 
under the BRS transformation is produced by adding the ghost term even after the gauge invariance 
of the Lagrangian is violated by the gauge fixing. It is, also, considered that 
the BRS invariance is not violated by the SSB in order for the theory to be renormalizable and for the $S$-matrix 
to be independent of the gauge parameter\cite{15}.

If we start with the gauge invariant Lagrangian containing the fields of 
the chiral transformation, the anomalies associated with the fermion triangle diagram on 
three currents must also disappear in order to guarantee the renormalizability of the theory 
and then the BRS invariance is ensured as is easily seen for our case (ii). 
Of course, it is evident that even after the SSB the above conditions for the theory shoud not be violated 
in order for the theory to be renormalizable. That is, the anomaly coefficients after the SSB must vanish 
and then the invariance under the BRS transformation must hold. However, in some cases, the invariance 
under the BRS transformation holds thought the anomaly coefficients do not vanish as in the case (i-b) 
in \S\ref{tri}. Therefore, in order for the theory to be renormalizable the anomaly coefficients must vanish 
and then also the BRS invariance will be ensured. This is because the BRS invariance is necessary for 
the unitarity of the $S$ matrix\cite{15}.  

As discussed in \S\ref{fer} and \S\ref{tri}, the cases (i-a) and (i-b) for the singlet lead to non-zero results 
for the anomaly coefficients though in the case (i-b) the total Lagrangian is invariant under 
the BRS transformation in contrast to the case (i-a) in which case the BRS transformation can not be 
defined as noted below $(\ref{25})$. However, in the case (ii) for the singlet the anomaly coefficients 
become zero for all possible triangle diagrams and the BRS invariance holds as is easily seen. 
Thus, it may be concluded that the possibility of the singlet $``1"$, i.e., the case (i), must be excluded for 
the theory to be renormalizable. We will show the invariance under the BRS transformation in the case (ii) for 
the counterparts of $``3"$ and $``\bar 3"$ briefly below. 

It is easily seen that the Lagrangian ${\cal L}_{gf}+{\cal L}_{gh}$ of the gauge fixing and 
the ghost terms in $(\ref{41})$ and $(\ref{42})$ is invariant under the following 
BRS transformations of the fields
\begin{eqnarray}
&&
\delta_BC^a=\frac{i}{2}\lambda f_{abc}C^bC^c,\nonumber\\
&&
\delta_B\overline{C^a} =\lambda\xi(\partial^\mu V^a_\mu-\frac {ig_a}{\xi}
\langle v|L^{\Phi a}|\tilde \Phi\rangle),\nonumber\\
&&
\delta_B\tilde{\Phi}=\lambda L^{\Phi a}(v+\tilde \Phi )C^a, \nonumber\\
&&
\delta_BV^a_\mu=-i\lambda D_\mu C^a ,\label{81}
\end{eqnarray}
where $D_\mu C^a\equiv\partial_\mu C^a +gf_{abc}V^b_\mu C^c$ and the BRS transformation is defined 
as usual, i.e., the substitution of the parameters in the gauge transformation 
$\beta^a \rightarrow i\lambda C^a$ with a constant hermitian Grassmann quantity $\lambda$. 
It is noted that the following equation of motion holds
\begin{eqnarray}
\partial^\mu D_\mu C^a+\frac{g_ag_b}{\xi}
\langle v|L^{\Phi a}L^{\Phi b}|v+\tilde \Phi\rangle C^b =0.\label{82}
\end{eqnarray}
where no sum over $a$ but the sum over $b$ is taken from 1 to 9.
The BRS transformation for the fermions is given by the above substitution of the parameters in the formulas 
such as the transformation in $(\ref{314a})$. It thus follows that if the Lagrangians except for the gauge fixing 
and the ghost terms are invariant under the BRS transformation even after the SSB, the total Lagrangian is 
also invariant under the transformation. 

It is evident that the original Lagrangians $(\ref{26a}),(\ref{26b})$,
$(\ref{26c})$ and $(\ref{210})$ are invariant under the BRS transformation given by $v=0$ in $(\ref{81})$ 
and even in the case of $v\neq 0$ for the leptons and the scalars because as noted in \S\ref{ssb} 
the Lagrangians and the fields in terms of the mass eigenstates are rewritten in the same form corresponding to 
those in terms of the weak interaction bases and the BRS transformation for the mass eigenstates is given in 
the same form as that for the weak interaction bases as noted in \S\ref{for} and \S\ref{ssb}. 

However, in the case (i-a) the Lagrangian for the quarks can not be rewritten with 
the similar form in terms of the weak interaction bases and the mass eigenstate bases also as 
noted in \S\ref{ssb} and furthermore the variation for the quarks $Q_{aL}$ can not be expressed in 
the form corresponding to those in terms of the weak interaction bases because the quarks in 
the different representations are mixed. In the case (i-b) as mentioned in \S\ref{fer} 
the Lagrangian for the quarks can be expressed in the same form corresponding to that 
in the weak interaction bases, but the propagators of the quarks are given for a set of 
the quarks belonging to the different representations (see $(\ref{52})$). Therefore, in this case 
the total Lagrangian is inariant under the BRS transformation but the anomaly coefficients 
are not zero in general.

In the case (ii) by taking into account the transformation property of the right-handed singlet 
of the quarks there is no problem in expressing the Lagrangian in terms of the weak interaction 
bases in the corresponding form in terms of the mass eigenstates as seen up to now. Thus, even after 
the SSB the anomalies as well as the FCNC disappear and the theory will become renormalizable. 
Of course an invariance under the BRS transformation is ensured.
\section{Discussion}\label{dis}

We discussed the $331$ model in which for the left-handed quark sectors in the three families 
the complex conjugate representations are assigned to the first and the second families and 
the fundamental representation is assigned to the third family, while the singlet assignment to 
the right-handed quarks associated with the left-handed quarks has the case of $``1"$ 
independent of the left-handed $3$ and $\bar 3$ and the case of $``1_3"$ and $``1_{\bar 3}"$ 
depending on the left-handed $``3"$ and $``\bar 3"$. Though the choices have an effect to 
the form of the Yukawa interactions of the quarks with the scalar fields as seen in $(\ref{210})$, 
the theory will be anomaly free and renormalizable at the stage of the original Lagrangian 
before the SSB. In this case, it is easily seen that an invariance under the BRS transformation 
holds after fixing to the $R_\xi$ gauge in addition to the ghost terms.

However, the situation after the SSB becomes quite different from that before the SSB. 
This is because except for the $J$ quarks the Yukawa interaction giving mass to the quarks through 
the SSB is different for both cases of the singlet. As shown in the previous sections, 
the anomaly coefficients can not be zero in general in the case of the adoption of the singlet 
$``1"$ independent of the left-handed $``3"$ and $``\bar 3"$ but in the case (ii) of 
the right-handed singlet $``1_3"$ associated with the fundamental representation and 
$``1_{\bar 3}"$ with the complex conjugate representation there appear no anomalies even after 
the SSB. It seems natural in the gauge theory with the chiral trasformation for us to take 
into account the transformation propperties of the singlet in physics but not mathematics 
as discussed in \S\ref{int}.

Though an invariance under the BRS trasformation holds in the case (i-b), the anomaly remains 
without vanishing as pointed out in \S\ref{brs} and thus the renormalizability of the theory 
is destroyed after the SSB. However, in the case (ii) of the singlet $``1_3"$ and $``1_{\bar 3}"$ 
the invariance under the BRS trasformation as well as disappearance of any anomaly holds even 
after the SSB. We may thus conclude that in construction of the model based on 
the chiral gauge theory with the fundamental and complex conjugate representations in 
the assignment of the representations to the basic particles in the families such as 
the $331$ model the counterparts of the fundamental and the complex conjugate representations 
should be subjected to the definite transformation according to that of each partner.
\appendix
\def\theequation{A.\arabic{equation}}
\section*{Appendix.1}

As is well known, the singlet appears in the following ways 
in the representation of ${\rm SU(3)}$
\begin{eqnarray*}
&&
({\rm i})\ 3\otimes 3 = 1\oplus 8,\\
&&
({\rm ii})\ 3\otimes 3\otimes 3 = 1\oplus 8\oplus 8\oplus 10,\\
&&
({\rm iii})\ {\bar 3}\otimes{\bar 3}\otimes{\bar 3}= 1\oplus 8\oplus 8\oplus\overline{10},
\end{eqnarray*}
where $3$ denotes the fundamental representation and $\bar 3$ 
the complex conjugate representation of ${\rm SU(3)}$. 
Of course, it is well known that the expressions from the left side for the singlet 
$``1"$ on the right side are equivalent mathematically to each other in the sense 
that the relation such as $3\otimes{\bar 3}={\bar 3}\oplus 6$ holds and 
then one of the configurations on the right-side given from 
the left side is rewritten to another singlet  configuration $``1"$. 
That is, it may be said that these three singlets are equivalent to each other under 
a transformation of ${\rm SU(3)}$. However, it is physically natural to consider 
their physical content different because if $3$ is assigned to a member of 
the fundamental fermion and each configuration on the right side is accepted as 
the quantity of a definite particle, the particle corresponding to the singlet in (i), 
called bosonic singlet, will express a boson-like one such as in the boson 
in the quark model but not a fermion or anti-fermion, and those in (ii) and (iii), 
called fermionic and antifermionic singlets, will express only a fermion and an anti-fermion, 
respectively. In other words, the bosonic singlet in (i) may be said the scalar 
with respect to both $3$ and /or $\bar 3$ or independent of 3 and $\bar 3$ 
in the sense that the configuration consists of the fundamental fermion and antifermion, 
but those in (ii) and (iii) are the scalar only with respect to the $3$ and $\bar 3$ 
in the sense that each of them consists of only the fermion and antifermion, respectively. 
Thus, it is physically natural to consider that there exist three kinds of 
the singlet corresponding to (i), (ii) and (iii). 
Of course, it does not need considering the distinction of these in almost all models 
except for the models such as the $331$ model because only 
the fundamental (or complex conjugate) representation for the lefdt-handed basic particles 
in the first stage is adopted and then the singlet of the counterparts is meant 
as in (ii) (or (iii)), while in the $331$ model both of the fundamental and 
complex conjugate representations are used to assignment of the representations and 
then the transformation property of the singlet affects the forms of the Yukawa interactions. 
It is, however, noted that the fermions in the families are not considered composite 
such as above (i), (ii) and (iii) and then the assignment of the representation to 
these fermions are free a priori provided that it is physically allowed and leads to 
a consistent result. 
\def\theequation{B.\arabic{equation}}
\section*{Appendix.2}
 
Let us consider a group ${\rm G}$ with the generators  $T^j \ (j=1,2,...,{\rm N})$ satisfying 
the commutation relations
\begin{eqnarray}
[T^j,T^k]=ic_{jkl}T^l,\label{b1}
\end{eqnarray}
where the generators are hermitian. We consider n complex scalar fields 
$\Phi_j \ ( j=1,2,...,{\rm n})$, arranged in a column vector $\Phi$ transforming according to 
the ${\rm n\times n}$ representation matrices $L^j$, 
which satisfy the same commutation relations as those in (B.1) and are normalized by 
${\rm tr }(L^jL^k) =t(L)\delta_{jk}$ in each irreducible component of $L^j$. 
The $\Phi$ is subject to the following relation under an infinitesimal transformation:
\begin{eqnarray}
\Phi^\prime_j =\Phi_j -i({\sl u}\cdot {\sl L})_{jk}\Phi _k.\label{b2}
\end{eqnarray}
Instead of n complex scalar fields $\Phi_j\ ( j=1,2,...,{\rm n})$, 
it is often convenient to use 2n hermitian scalar fields by writing 
$\Phi_j=(\Phi_{2,j}+i\Phi_{1,j})/\sqrt 2$ with hermitian $\Phi_{1,j}$ and $\Phi_{2,j}$. 
Then, due to the relation ${\rm U(1)\sim SO(2)}$ the transformation (B.2) is rewritten 
in terms of a direct product as follows:
\begin{eqnarray}
{\hat \Phi}^\prime
&&=\left(I-iI_2\otimes{\sl u}\cdot\left({\sl L}-{\sl L}^T\right)/2
-i\tau_2\otimes{\sl u}\cdot\left({\sl L}+{\sl L}^T\right)/2\right){\hat\Phi},\label{b3}
\end{eqnarray}
where $I_2$ and $\tau_2$ are $2\times 2$ unit and Pauli matrices with elements 
$(\tau_2)_{\alpha\beta}=-i\epsilon_{\alpha\beta}=i\epsilon_{\beta \alpha},\epsilon_{12}=1$, 
and ${\hat\Phi}=(\Phi_{1,1},\Phi_{2,1},\Phi_{1,2},\Phi_{2,2},\cdots,\Phi_{1,{\rm n}},\Phi_{2,{\rm n}})^T$. 
An action of $I_2 \otimes L^j$ and $\tau_2 \otimes L^j$ on $\hat \Phi $ is given by 
$(I_2\otimes L^j\hat\Phi)_{\alpha,k}={\delta}_{\alpha\beta}(L^j)_{kl}\hat\Phi_{\beta,l}$ 
and $(\tau_2\otimes L^j\hat\Phi)_{\alpha,k}=
{(\tau_2)}_{\alpha\beta}(L^j)_{kl}\hat\Phi_{\beta,l}$, respectively.  
It then follows that the representation matrices $I_2\otimes(L^j-{L^j}^T)$ 
and $\tau_2\otimes (L^j+{L^j}^T)$ are hermitian and antisymmetric: 
${(I_2\otimes(L^j-{L^j}^T)})^T=-I_2\otimes(L^j-{L^j}^T)$ and 
${(\tau_2\otimes(L^j+{L^j}^T)})^T=-\tau_2\otimes(L^j+{L^j}^T)$. 
If the covariant derivative is written in the form $D_\mu\Phi\equiv(\partial_\mu-iM_\mu)\Phi$ 
with hermitian $M_\mu$, then the corresponding covariant derivative 
$\hat D$ for the $\hat\Phi$ becomes 
$\hat D_\mu \hat\Phi=(\partial_\mu-i\{I_2\otimes(M_\mu-M^T_\mu )/2
+\tau_2 \otimes (M_\mu +M^T_\mu )/2\})\hat \Phi $.

It is seen that the following relations hold
\begin{eqnarray}
{\Psi}^\dag{\Phi}
&&=\frac{1}{2}{\hat\Psi}^T(I_2+\tau_2)\otimes I_n{\hat\Phi},\nonumber\\
{\Psi}^\dag{\sl u}\cdot {\sl L}{\Phi}
&&=\frac{1}{2}{\hat\Psi}^T(I_2+\tau_2)\otimes{\sl u}\cdot{\sl L}{\hat\Phi},\nonumber\\
&&\hspace{-1cm}=\frac{1}{2}{\hat\Psi}^T\{(I_2+\tau_2)\otimes I_n\}\{I_2\otimes {\sl u}\cdot ({\sl L}-{\sl L}^T)/2
+\tau_2\otimes{\sl u}\cdot({\sl L}+{\sl L}^T)/2)\}{\hat\Phi}\label{b4}\\
(D^\mu\Psi)^\dag D_\mu\Phi
&&=\frac{1}{2}(\hat D^\mu\hat\Psi)^T(I_2+\tau_2)\otimes I_n \hat D_\mu\hat\Phi\nonumber\\
&&\hspace{-1cm}=\frac{1}{2}(\partial^\mu\hat\Psi^T I_2\otimes I_n+i\hat\Psi^TI_2\otimes M^\mu)
\{(I_2+\tau_2)\otimes I_n\}I_2\otimes(I_n\partial_\mu-iM_\mu)\hat\Phi,\nonumber
\end{eqnarray}
where $I_n$ is ${\rm n\times n}$ unit matrix.@  

It is often useful to consider anti-commutators which are hermitian 
but have non-zero trace and may be written as follows
\begin{eqnarray*}
\{L^j,L^k\}=d_{jkl}L^l+\frac{2{\rm tr}(L)}{n}\delta_{jk}I_n,
\end{eqnarray*}
where $d_{jkl}$ are totally symmetric in three indices and given by 
$d_{jkl}={\rm tr}(\{L^j,L^k\}L^l)/{{\rm tr}(L)}$. 
It is known that the non-zero structure constants $c_{jkl}$ and 
the symmetric coefficients $d_{jkl}$ are given in the case of
 the $3\times 3 $ Gell-Mann matrices $L^j =\lambda^j/2 \ (j=1,2,\cdots 8)$ as follows:
\begin{eqnarray*}
&&
c_{123}=1,\\
&&
c_{147}= c_{165}= c_{246}= c_{257}= c_{345}= c_{376}={1}/{2},\\
&&
c_{458}= c_{678}=\sqrt{3}/{2},\\
&&
d_{146}=d_{157}=d_{246}=-d_{247}=d_{256}=d_{344}=d_{355}=-d_{366}=-d_{377}=1,\\
&&
d_{118}=d_{228}=d_{338}=-d_{888}=2\sqrt{3}/{3},\\
&&
d_{448}=d_{558}=d_{668}=d_{778}=-\sqrt{3}/{3}.
\end{eqnarray*}

In the case of ${\rm SU(n)}$, the ${\rm (n^2-1)}$ matrices $L^j$ of the fundamental representation 
can be separated in two classes, symmetric matrices ${L^j}^T =L^j$ and antisymmetric matrices ${L^j}^T =-L^j$.  
A set of numbers satisfying the former relation is put ${\rm F_1}$ with ${\rm (n-1)(n+2)/2}$ numbers and that of 
the latter ${\rm F_2}$ with ${\rm n(n-1)/2}$ numbers. Then it follows that the representation matrices 
$\{(I_2\otimes (L^j-{L^j}^T)/2,\tau_2\otimes (L^k+{L^k}^T)/2\}$, which are antisymmetric and are equivalent to 
the ${\rm (n^2-1)}$ matrices contained in the ${\rm n(2n-1)}$ represenation matrices of the group ${\rm SO(2n)}$, 
become $\{I_2\otimes L^j,\tau_2\otimes L^k \}$ for $j\in {\rm F_2}$ and $k\in {\rm F_1}$. Thus, they satisfy 
the same commutation  relation as those for $L^j$, because the relations 
$[ A\otimes B, C\otimes D]=[A,C]\otimes BD+CA\otimes [B, D]$ holds and the first term vanishes. 
However, the anti commutators $\{ A\otimes B, C\otimes D\}$ can not be expressed in terms of the above 
antisymmetric ones because these commutators are symmetric matrices. 
In the case of ${\rm SU(2)}$ and ${\rm SU(3)}$, ${\rm F_1}$ consists of the matrices with the numbers 
$\{ 1, 3\}$ and $\{ 1, 3, 4, 6, 8\}$ and ${\rm F_2}$ with $\{2\}$ and $\{2, 5, 7\}$, respectively.

\end{document}